\documentclass[12pt]{iopart}
\usepackage{graphicx}
\usepackage[table]{xcolor}
\usepackage{adjustbox}
\usepackage{algorithm}
\usepackage{algorithmic}
\usepackage{hyperref}

\begin{document}

\title{Machine learning-based prediction of $Q$-voter model in complex networks}

\author{Aruane M. Pineda}
\address{Institute of Mathematical and Computer Sciences, University of São Paulo, São Carlos, São Paulo, Brazil}
\address{Mathematics Institute, University of Warwick, Coventry, England, UK }
\ead{aruane.pineda@usp.br}

\author{Paul Kent}
\address{Mathematics Institute, University of Warwick, Coventry, England, UK}
\ead{Paul.Kent@warwick.ac.uk}

\author{Colm Connaughton}
\address{Mathematics Institute, University of Warwick, Coventry, England, UK}
\address{London Mathematical Laboratory, London, England, UK}
\ead{c.connaughton@lml.org.uk}

\author{Francisco A. Rodrigues}
\address{Institute of Mathematical and Computer Sciences, University of São Paulo, São Carlos, São Paulo, Brazil}
\ead{francisco@icmc.usp.br}

\begin{abstract}

In this article, we consider machine learning algorithms to accurately predict two variables associated with the $Q$-voter model in complex networks, i.e., (i) the consensus time and (ii) the frequency of opinion changes. Leveraging nine topological measures of the underlying networks, we verify that the clustering coefficient (C) and information centrality (IC) emerge as the most important predictors for these outcomes. Notably, the machine learning algorithms demonstrate accuracy across three distinct initialization methods of the $Q$-voter model, including random selection and the involvement of high- and low-degree agents with positive opinions. By unraveling the intricate interplay between network structure and dynamics, this research sheds light on the underlying mechanisms responsible for polarization effects and other dynamic patterns in social systems. Adopting a holistic approach that comprehends the complexity of network systems, this study offers insights into the intricate dynamics associated with polarization effects and paves the way for investigating the structure and dynamics of complex systems through modern methods of machine learning.

\end{abstract}

\vspace{2pc}

\section{Introduction}

Interactions among the components of a complex system have given rise to properties not present in its isolated parts \cite{Thurner18}. For instance, the collective behavior of ants in a colony provides a compelling illustration of emergence. While individually following simple rules, ants exhibit complex behaviors such as efficient food foraging, elaborate nest construction, and coordinated defense \cite{Boccara10}. Such an emergence phenomenon significantly extends beyond the natural world, since it also manifests within our society through intricate interactions among agents, groups, and institutions.

A substantial consequence of emergence is social polarization, according to which agents develop increasingly extreme opinions and display diminished tolerance for opposing viewpoints, ultimately leading to societal divisions. Numerous studies have associated the phenomenon with negative outcomes in political contexts, as seen in the recent elections in both Brazil and the United States \cite{del2016spreading, flaxman2016filter, barbera2015tweeting, Conover2011}. In Brazil, heightened polarization culminated in a significant event on January 8, 2023, when key institutions in Brasília, the capital of Brazil, were invaded. This event was the result of escalating tensions stemming from polarized political discourse. The Supreme Federal Court, the National Congress building, and the Presidential Palace were among the targeted institutions. Similarly, the United States also faced its own challenges associated with polarization. A notable incident occurred on January 6, 2021, when a crowd stormed the United States Capitol in an attempt to overturn the results of the presidential election. Therefore, the causes and effects of polarization in social networks must be comprehended so that effective communication strategies and social interventions that mitigate its detrimental impact can be designed \cite{bessi2016social, centola2018experimental}.

Towards a deeper understanding of social polarization, various mathematical models have been developed \cite{castellano2009statistical} and the most sophisticated ones have recently considered the dynamics of interactions between agents and their underlying structure. Indeed, consensus models must be simulated in complex networks to be more realistic, since the network topology heavily influences both their dynamics and the final result of consensus generation \cite{soares2021empirical}. 

Several models, including the Ising model, the Sznajd model, the voter model, the naming game, the bounded confidence model, and the $Q$-voter model, address complex phenomena stemming from interactions among individuals in social and physical contexts and are adapted for complex networks. Researchers employ these models to identify conditions fostering consensus emergence and network features facilitating the process. Simulations within complex networks are crucial to achieving a more accurate portrayal of consensus formation. The intricate network topology significantly influences model dynamics and consequently impacts consensus generation outcomes \cite{soares2021influence}. The Ising model, originating from physics, focuses on material magnetization by representing the magnetic orientation of spins in a three-dimensional lattice. The interaction between neighboring spins aims to minimize the system's energy, leading to phenomena like the Ising phase transition \cite{ising1925beitrag,stauffer2007ising,newman1999monte,kuperman2001small}. In contrast, the Sznajd model explores how similar opinions can influence others. The premise is that people with coinciding opinions are more likely to persuade others, leading to the formation of opinion clusters \cite{snajd2005,sznajd2000opinion,lima2007nonequilibrium, rodrigues2005surviving}.
Meanwhile, the voter model simplifies decision-making in a population, where individuals adopt the majority opinion of their neighbors, illustrating how social influences can drive convergence towards dominant opinions or polarization \cite{liggett1985interacting,castellano2009statistical}. The naming game addresses language evolution, where individuals attempt to communicate and reach a consensus on names for concepts, balancing communicative efficiency and linguistic diversity \cite{baronchelli2006sharp,castellano2009statistical, baronchelli2006mean,baronchelli2006naming}. Bounded confidence explores how opinions change through social interactions, assuming people update their opinions only when the difference from others' opinions falls within a specific limit \cite{deffuant2000mixing,hegselmann2002opinion}.
Finally, the $Q$-voter model offers an approach to simulate collective decisions within groups of individuals. In this model, each agent adopts the opinion of one of its randomly selected $Q$-neighbors. These $Q$-neighbors represent a subset of neighboring agents, and the extent of this subset, denoted by $Q$, significantly influences the dynamics of opinion diffusion. This model investigates how connectivity and information exchange between individuals impact consensus formation. By studying how opinions spread through the $Q$-voter framework, researchers gain insights into the emergence of consensus, polarization, or the coexistence of diverse viewpoints within a population \cite{castellano2009statistical,malarz2009analysis,pinheiro2015asymmetric}. In \cite{doniec2022consensus}, researchers investigate the impact of polarization in the three-state $Q$-voter model, considering limited confidence and noise. By incorporating these factors, the study reveals how agent interactions lead to the formation of groups with divergent opinions, complicating the convergence to a single opinion. Similarly, \cite{lipiecki2022polarization} examines the role of anticonformity and limited confidence in the $Q$-voter model. This study demonstrates that anticonformity amplifies polarization and emphasizes the coexistence of groups with similar yet distinct opinions, especially when limited confidence is present. Furthermore, \cite{krueger2017conformity} introduces a mathematical model that examines the effects of conformity and anticonformity on opinion polarization. This study investigates a similar opinion dynamics model based on the $Q$-voter, analyzing how the interplay between these behaviors influences the formation of groups with divergent opinions. These collective studies substantially contribute to a deeper understanding of the underlying dynamics of opinion polarization in social contexts.

Empirical investigations have consistently provided compelling evidence that different network topologies exhibit varying degrees of polarization and consensus formation 
\cite{soares2021modular,fernandez2014effect,de2013effects,liu2020effect, brugnano2018role}. For example, recent studies have shown the adoption of the $Q$-voter model within modular networks can result in highly polarized public opinions \cite{soares2021modular}. On the other hand, in scale-free networks, highly connected agents can expedite the process of consensus formation while potentially amplifying extreme polarization \cite{fernandez2014effect}. Furthermore, studies have delved into the influence of network clustering, degree distribution, and other network properties on the dynamics of consensus formation \cite{castellano2009statistical}, highlighting the crucial role of network topology in the development of realistic models for comprehending consensus formation within complex networks. By considering the intricate interplay between network structure and opinion dynamics, researchers can attain a more comprehensive understanding of the factors that shape the emergence of consensus and polarization.

Given the significant influence of network topology on the emergence of consensus, an essential question is whether it is feasible to develop a machine learning model that can forecast dynamic variables based on network properties. Such an inquiry has been widely explored in various fields, including the prediction of both epidemics in human contact networks \cite{keeling2005networks,rodrigues2019machine} and synchronization in coupled oscillators \cite{pecora1998synchronization,rodrigues2019machine}. The investigations have not only demonstrated the possibility of forecasting the behavior of dynamic systems from the network topology but also underscored the importance of comprehending the relationship between network structure and dynamics in those systems as in the recent article by Brooks and Porter \cite{brooks2020model}. Both our study and the one by Brooks and Porter share a common focus on complex phenomena within social networks. We employ interdisciplinary approaches that integrate complex network theory, system dynamics, and machine learning. Both studies acknowledge the pivotal role of network structure in shaping social dynamics and investigate opinion dynamics within social networks, although with different emphases. While both studies delve into opinion dynamics, our research primarily centers on utilizing machine learning to predict variables based on the $Q$-voter model. In contrast, Brooks and Porter's research delves into how media exposure influences ideological content within social networks. This distinction underscores the significance of media in their study, while our research places a strong emphasis on network structure and agent interactions.

The application of machine learning algorithms in the study for the prediction of consensus time and frequency of opinion changes in the $Q$-voter model offers several advantages. Machine learning promotes the capture of complex patterns, learning from historical data, and adaptation to evolving dynamics; it is a powerful tool for uncovering intricate relationships and enhancing predictive accuracy. Moreover, its use in the context of $Q$-voter represents a novel approach, pushing the boundaries of traditional analysis and providing new insights into the mechanisms driving opinion dynamics in complex social systems. The consensus is a significant metric that indicates the level of agreement among agents in a network. Conversely, the frequency of opinion changes reflects a network’s ability to maintain its beliefs and showcases the level of volatility in the system. Understanding and mitigating the effects of polarization in complex network systems is of utmost importance, as it can significantly impact both the consensus formation process and the stability of opinions within a network. Both metrics play a crucial role in the comprehension of the behavior of social systems and offer insights into the factors contributing to stability or instability within such systems \cite{moretti2013mean}.  

This study provides valuable insights into the intricate relationship between network structure and social dynamics, highlighting the potential of complex network measures for analyzing dynamic systems. Additionally, it demonstrates the effectiveness of complex network structures in accurately predicting the consensus time and frequency of opinion changes within the $Q$-voter model using machine learning algorithms. The significance of each network feature in these predictions was evaluated, revealing the clustering coefficient (C) and information centrality (IC) as the most influential measures for predicting these outcomes. Furthermore, the robustness of these predictions was tested using three distinct initialization methods in the $Q$-voter model, specifically assessing the model's behavior when initialized with high degree, low degree, and a random selection of agents with positive opinions.

The article is organized as follows: Section \ref{Methods} is divided into four parts. The first part introduces the simulated $Q$-voter model, Subsection \ref{networks} describes the investigated networks, Subsection \ref{measures} explains the network measurements, and Subsection \ref{models} presents the machine learning algorithms used for prediction. Section \ref{results} provides the results, and Section \ref{conclusions} is dedicated to relevant observations and conclusions.

\section{Methods}
\label{Methods}

\subsection{Stochastic simulation of $Q$-voter model}

In the context of the $Q$-voter model, a group of $Q$ agents ($Q$-voters) influences the opinion of a single agent. This interaction determines the number of neighbors considered by an agent for decision-making, as dictated by the parameter $Q$. This model is particularly interesting for studies of social dynamics since it captures the impact of group influence, conformity, and social reinforcement on opinion dynamics. Furthermore, it exhibits a rich phase-transition behavior, depending on the value of $Q$ and network topology, leading to various outcomes such as consensus, fragmentation, and coexistence of opinions \cite{lorenz2007continuous, moretti2013griffiths, guerra2004information, mobilia2003single, galam2002minority}. Introduced in \cite{castellano2009nonlinear}, its applicability extends to all integer-values of $Q$ $>$0, meaning that $Q$ can encompass a range of values greater than 0. Furthermore, by setting $Q$=1, we directly return to the standard voter model. Within this framework, the possibility of repetition is considered, implying that a specific neighbor can be selected multiple times. Thus, when $Q$ is greater than the number of neighbors (the degree of a node), the opinion of the same neighbor will be taken into account more than once.

Consider a network of $N$ voters (also known as agents, nodes, spins, or individuals). Each is defined by a single dynamical binary variable $s(x, t)=j$, where $j=+1$ or $j=-1$, $x=1,..., N$, and $t$ represents time. From a social standpoint, $s(x, t)$ represents a two-point psychometric scale (yes/no, agree/disagree) opinion of an agent placed at node $x$ at time $t$ on a particular subject.

The initial fraction of agents with positive opinions ($p+$) is fixed at the beginning of the simulation and randomly distributed to the network nodes. Parameter $\epsilon$ represents the probability of an agent $x$ acting independently of their neighbors, indicating their unwillingness to yield to group pressure. Consequently, ($1-\epsilon$) represents conformity, influencing the likelihood of an agent adopting the majority opinion of her/his $Q$ neighbors. Note that the individual opinion of the selected agent $x$ is not taken into account in the probability of opinion change or retention in the dynamics. Table \ref{Q-voter-parameters} shows the fixed parameters of $Q$-voter, including the number of nodes in complex networks ($N = 1, 000$), probability of an agent acting independently ($\epsilon = 0.01$), an initial fraction of agents with positive opinions ($p+ = 0.20$), and the number of neighbors ($Q = 2$). The value of $\beta$ represents the probability of an agent changing their opinion to the opposite when there is no consensus among their neighbors. 

The parameters were fixed toward establishing a consistent baseline for our machine learning-based prediction of consensus time and frequency of opinion changes. Consensus time is the relaxation time of a finite-size system needed to approach a stationary state. By keeping them constant, our exploration can focus on the impact of other variables and a more thorough analysis of our machine learning model’s predictive performance regarding the desired outcomes can be conducted. The initial percentage of agents selected was modified to have a positive opinion in three ways: through random agent selection, and by selecting high- and low-degree agents.

\begin{table}[hb]
    \centering
    \caption{$Q$-voter model parameters with default values.}
    \label{Q-voter-parameters}
    \small 
    \setlength{\tabcolsep}{6pt} 
    \begin{tabular}{p{2cm}cc} 
    \hline
         Parameter & Default Value & Description \\
         \hline
         $N$ & 1,000 & Number of nodes  \\
         $\epsilon$ & 0.01 & Probability of an agent acting independently (non-conformity) \\
         $Q$ & 2 & Neighbor consideration for decision-making  \\
         $p_+$ & 0.20 & Initial fraction of agents with positive opinions \\
         $\beta$  & 0.20 & Probability to alter opinion with no consensus among neighbors \\
         \hline
    \end{tabular}
\end{table}

\begin{figure}[htbp]
\centering
\includegraphics[width=0.90\textwidth]{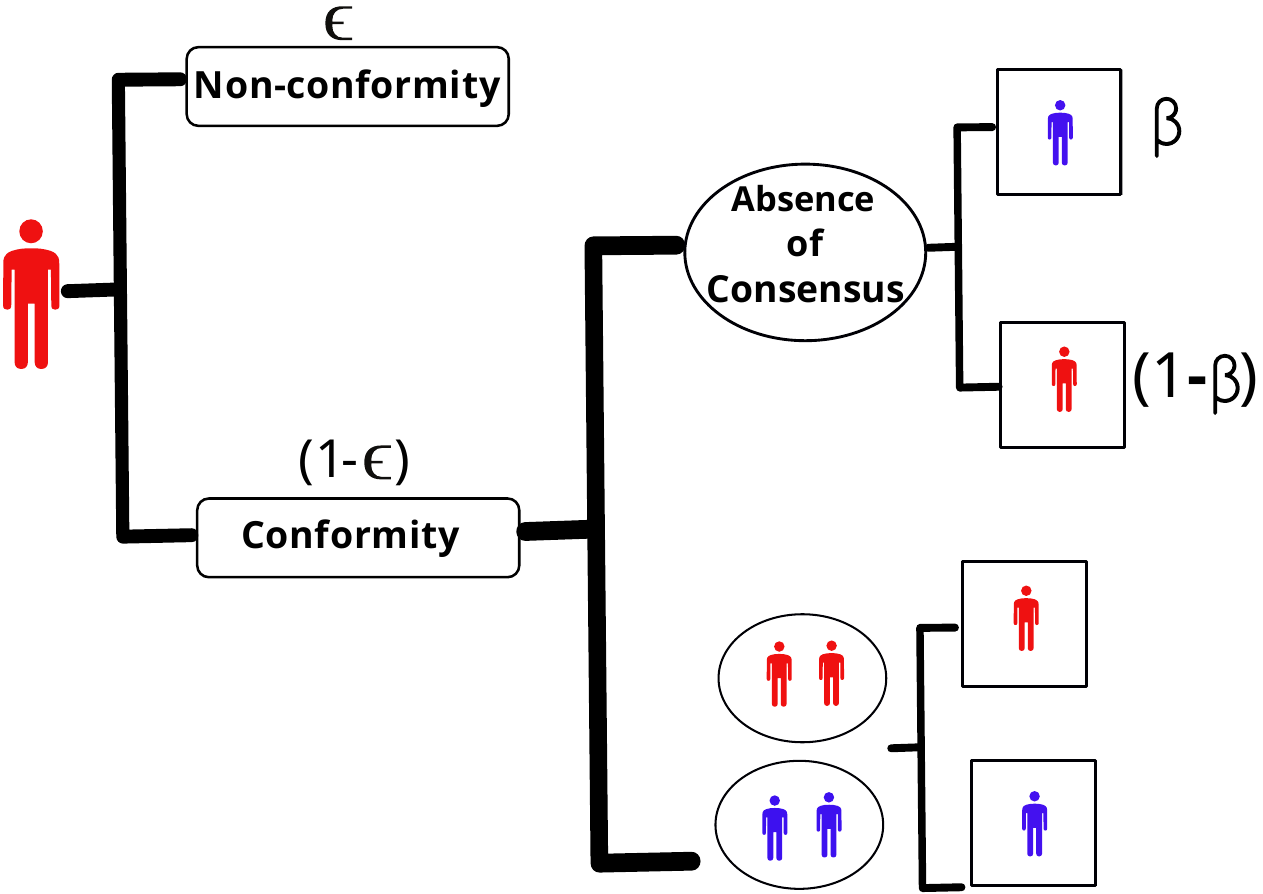}
\caption{Illustration of the $Q$-voter model: All agents have a binary opinion, represented here by the colors red and blue. Suppose an agent has a red opinion; then, their opinion can be altered based on the following social response: the probability of non-conformity, i.e., reluctance to yield to group pressure, with a probability $\epsilon$ of changing their opinion. Alternatively, there is conformity (1-$\epsilon$), which represents the probability of acting like their neighbors. If the neighbors have a consensus, meaning they all share the same opinion, the agent will switch to the blue color or remain in the red color. However, if there is no consensus among the neighbors, with a probability $\beta$, the agent will switch to the blue color, and with a probability of 1-$\beta$, they will maintain their opinion. The figure was created by the authors and is based on \cite{realqvoter}.}
\label{qvoter_illustration}
\end{figure}

Algorithm 1 exemplifies the stochastic simulation, followed by an illustration in Figure \ref{qvoter_illustration} of the model. In other words, all agents have a binary opinion, represented here by the colors red and blue. Suppose an agent has a red opinion; then, their opinion can be altered based on the following social response: the probability of non-conformity, i.e., reluctance to yield to group pressure, with a probability $\epsilon$, of changing their opinion. Alternatively, conformity (1-$\epsilon$) represents the probability of behaving like their neighbors. If the neighbors share a consensus, meaning they all have the same opinion, the agent will switch to the blue color or remain in the red color. However, if there is no consensus among the neighbors, with a probability $\beta$, the agent will switch to the blue color, and with a probability of 1-$\beta$, they will maintain their opinion.

\begin{algorithm}
\caption{$Q$-voter model algorithm}
\begin{algorithmic}[1]
\STATE Initialize a complex network of size $N$ representing the agents
\STATE Assign each agent a binary variable, $s(x,t)$ with $x \in [1,N]$ at time 
$t$, whose values +1 or -1 representing two opposing opinions ($j=+1$ or $j=-1$)
\FOR{each time step $t$}
    \STATE Randomly select an agent $x$
    \STATE Randomly choose $Q$ neighbors of agent $x$ (allowing for repetition)
    \IF{all $Q$ neighbors have the same state}
        \STATE agent $x$ takes the value of the $Q$ neighbors
    \ELSE
        \STATE agent $x$ flips with probability $\epsilon$
    \ENDIF
    \STATE Update the time
\ENDFOR
\end{algorithmic}
\end{algorithm}

\subsection{Networks}
\label{networks}

Nine complex network measures were examined, as discussed in Subsection \ref{measures}. The analysis involved eight distinct topological structures, including Erdős–Rényi \cite{erdHos1960evolution}, Barabási–Albert linear \cite{barabasi1999emergence}, Barabási–Albert nonlinear with $\alpha=0.5$ and $\alpha=1.5$ \cite{onody2004nonlinear}, Lancichinetti-Fortunato-Radicchi (LFR) graphs \cite{lancichinetti2008benchmark}, Watts–Strogatz \cite{watts1998collective}, Waxman \cite{waxman1988routing}, and path graph \cite{hagberg2014networkx}. The Erdős-Rényi network model is generated by randomly adding connections between nodes with a uniform probability. In contrast, the non-linear Barabási-Albert model is constructed iteratively, incorporating preferential attachment of new nodes to existing ones through a non-linear function that considers the node's connections. The LFR model is widely employed for creating networks with realistic community structures, assigning nodes to communities based on degree and community size distributions, and establishing connections that consider both intra- and inter-community links. The Watts-Strogatz model introduces the concept of small-world networks by randomly rewiring a portion of links in a regular lattice. A path graph is a specific type of graph in graph theory that consists of a linear sequence of connected nodes, where each node is linked to the next node in the sequence by a single edge. This creates a structure resembling a straight line of nodes, and it is often used as a simple representation of an ordered sequence of elements or events. A path graph is created by defining the nodes in the desired order and connecting them sequentially with edges. Lastly, the Waxman model takes into account geographic proximity and node attractiveness to determine the formation of connections, considering both physical distances and random appeal. For each of the mentioned networks, \ref{Construction of Complex Networks} provides details of the Python functions used, and  100 unique instances were generated, with each network consisting of $N = 1,000$ nodes and an average degree ranging from 9 to 10, that is, the available dataset comprises 800 instances of complex networks (denoted as $i$).

\subsection{Network Measurements}
\label{measures}

To effectively capture and explain the dataset's predominant variability, a visual representation of the principal component analysis (PCA) plot is provided in \ref{apendix:PCA}. This PCA plot serves as a powerful tool for gaining insights into the underlying patterns and structures within the dataset. Subsequently, the $Q$-voter model was simulated in each of these structures to measure the time taken to reach consensus ($Y_i$) and the total number of opinion changes that occurred in the model ($C_i$). It is hypothesized that both $Y_i$ and $C_i$ can be predicted using a feature vector derived from the network structure, denoted as $\mathbf{X_i} = {X_{i1}, X_{i2}, \ldots, X_{ik}}$, where $X_{ik}$ represents the $k$-th measure extracted from network $i$. The subsequent explanation primarily focuses on the prediction of $Y_i$, although the same process is applied to the prediction of $C_i$. Therefore, the learning model is defined by

\begin{equation}
Y_{i}= f(\mathbf{X}_{i}) + \delta.
\end{equation}

The goal is to infer the function $f()$ that relates $Y_i$ to the network measures. Estimating $Y_i$ is treated as a regression problem, where $\delta$ represents a random error term independent of $\mathbf{X}_i$, following a normal distribution with a mean of zero and a standard deviation of $\sigma$. While feature selection and model comparison algorithms can be used to identify components of $\mathbf{X}_i$ that contribute to predicting $Y_i$, this study employed conventional network measures which are presented in Table \ref{table_measures}.

The first measure utilized in this study was the clustering coefficient (C), a local measure, which quantifies the extent to which nodes in a network tend to form tightly connected clusters. It assesses the likelihood of two neighbors of a node being connected, reflecting the local clustering patterns within the network \cite{watts1998collective}. Closeness centrality (CLC), another local measure, was employed to calculate the proximity of a node to all other nodes in the network. It reflects the average distance between a node and all other nodes, indicating the efficiency of information or resource flow within the local neighborhood of a node \cite{freeman1979centrality}. Betweenness centrality (BC) is a measure that identifies nodes acting as critical intermediaries in the network. BC quantifies the extent to which a node lies on the shortest paths between other pairs of nodes, thus indicating its influence over the flow of information or resources within its vicinity \cite{freeman1977set}. The shortest path length (SPL) measures the minimum number of edges required to traverse between two nodes in the network, providing insights into network connectivity and the efficiency of information or resource transfer within local regions of the network \cite{newman2010networks}. Degree Pearson correlation coefficient (PC) examines the correlation between the degrees of connected nodes, capturing the tendency of nodes with similar degrees to connect and indicating the presence of assortativity or disassortativity within the network \cite{newman2018networks}. Information centrality (IC) assesses the importance of a node based on its ability to control the flow of information in the network, considering the number of shortest paths that pass through the node \cite{stephenson1989rethinking}. Subgraph centrality (SC) measures the importance of a node within its local subgraph by considering the closed walks that pass through the node, capturing its influence within specific network neighborhoods \cite{estrada2008communicability}. Approximate Current Flow Betweenness Centrality (AC) quantifies the extent to which a node controls the flow of electric current in the network, considering the current paths between all pairs of \cite{brandes2005modularity}. Finally, Eigenvector centrality (EC) determines a node's importance based on its neighboring nodes' centrality, assigning higher importance to nodes connected to other important nodes and capturing the concept of influence \cite{bonacich1987power}. Such measures, collectively used here, provide valuable insights into complex network structures, connectivity, efficiency, influence, and community organization \cite{costa2007characterization}. Details and equations for each of the mentioned measures, along with the Python functions used, are provided in \ref{Network Measurement Details}.

\begin{table}[hb]
\centering
\caption{Measures of complex networks used here.}
\begin{tabular}{|c|l|c|}
\hline
\rowcolor[HTML]{CCCCCC}
 & \textbf{Network Measures} & \textbf{Acronym} \\ \hline
\rowcolor[HTML]{FFFFFF}
$X_{1}$ & Clustering coefficient & C \\
\rowcolor[HTML]{FFFFFF}
$X_{2}$ & Closeness centrality & CLC \\
\rowcolor[HTML]{FFFFFF}
$X_{3}$ & Betweenness centrality & BC \\
\rowcolor[HTML]{FFFFFF}
$X_{4}$ & Shortest path length & SPL \\
\rowcolor[HTML]{FFFFFF}
$X_{5}$ & Degree Pearson correlation coefficient & PC \\
\rowcolor[HTML]{FFFFFF}
$X_{6}$ & Information centrality & IC \\
\rowcolor[HTML]{FFFFFF}
$X_{7}$ & Subgraph centrality & SC \\
\rowcolor[HTML]{FFFFFF}
$X_{8}$ & Approx. Current flow betweenness centrality & AC \\
\rowcolor[HTML]{FFFFFF}
$X_{9}$ & Eigenvector centrality & EC \\ \hline
\end{tabular}
\label{table_measures}
\end{table}

\subsection{Machine learning algorithms}
\label{models}

The machine learning algorithms utilized are the least absolute shrinkage and selection operator (LASSO), multi-layer perceptron regressor (MLP), random forest (RF), and extreme gradient boosting (XGBoost). Among the several techniques used to improve the proposed machine learning algorithms, nested cross-validation, shuffle, and grid search are highlighted. The former is a multi-round cross-validation procedure adopted in machine learning for model selection and performance assessment \cite{wainer2021nested}. It is a more rigorous model selection and performance evaluation than traditional cross-validation since it reduces the risk of overfitting and provides a more accurate estimate of the model’s performance on unseen data \cite{cawley2010over}. Its main idea is the existence of an outer loop, which divides the data into training and test sets, and an inner one, which uses cross-validation to determine optimal values for the model’s hyperparameters. Shuffle was employed during nested cross-validation to avoid possible biases in the selection of training and testing data, ensuring the model learned in a balanced way throughout the range of data. Finally, grid search searched for the best model hyperparameters by systematically exploring different combinations of possible values for them. The set of techniques used significantly contributed to the development of a more robust and accurate model. A 5-fold outer shuffle cross-validation and a 5-fold inner cross-validation were also adopted, following similar approaches described in previous studies \cite{leaver2018fronto}. During the inner folds, a grid search hyperparameter optimization was performed - specific details can be found in Table \ref{table-auti:gridsearch} in the \ref{apendix:grid}.

The coefficient of determination, R2, is a metric used to measure how well a regression model fits the data \cite{nakagawa2017coefficient}. However, when we add more predictors to the model, R2 can increase even if these new predictors don't really help explain the variation in the dependent variable. To address this, we use the R2 adjusted, which considers the number of predictors and penalizes the inclusion of irrelevant ones. This adjustment gives us a more accurate evaluation of how well our model predicts the outcome. In simpler terms, we prefer R2 adjusted over R2 because it prevents values from being artificially inflated by including unnecessary predictors. This ensures a more reliable assessment of our model's performance.

\begin{figure}[htbp]
\centering
\includegraphics[width=0.90\textwidth]{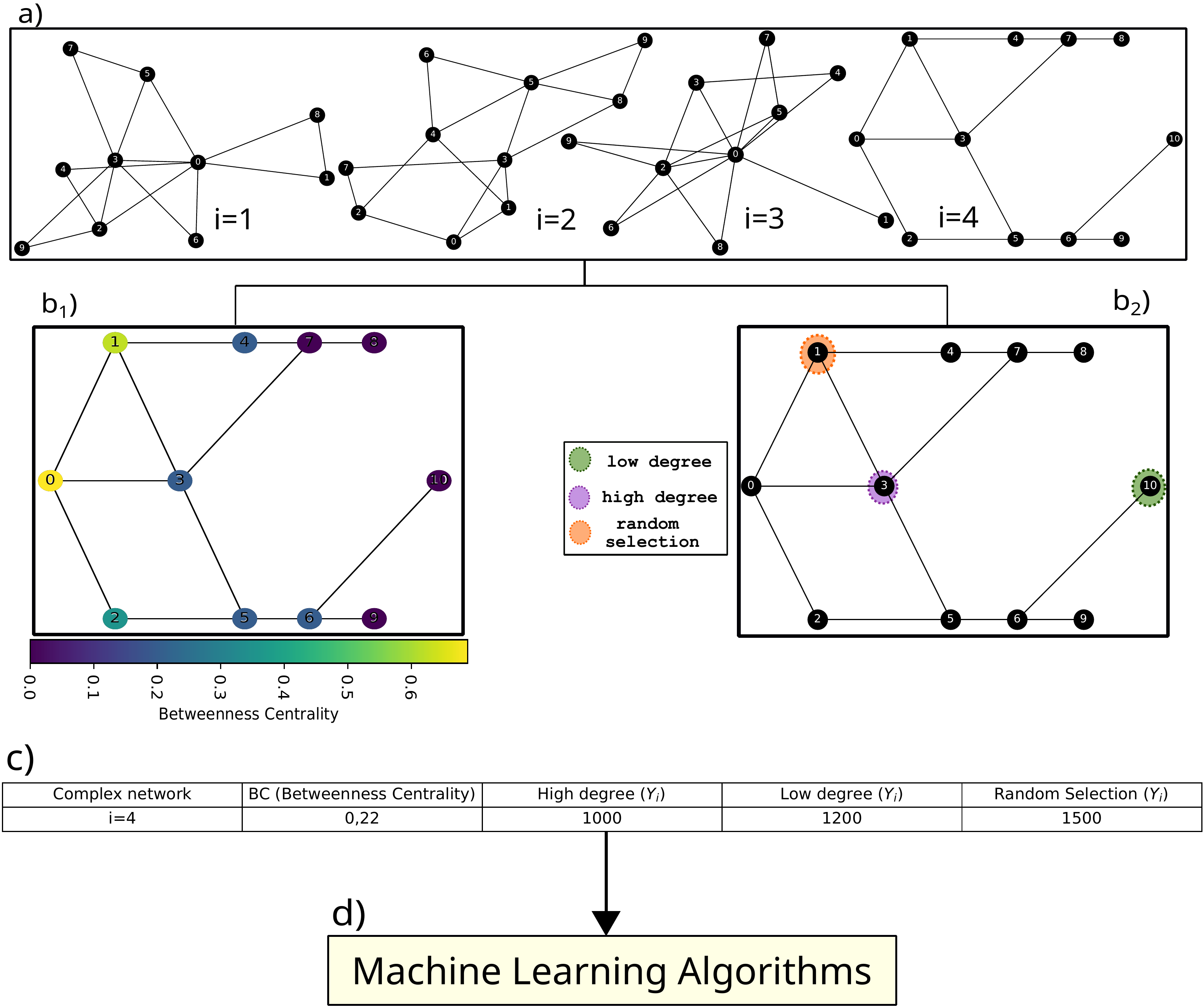}
\caption{Schematic Overview of the  process outlined in this article. The process involves several key steps: a) Generation of complex networks: In this initial step, we create complex networks for analysis. In this illustrative example, we generate four networks labeled as $i=1$, $i=2$, $i=3$, and $i=4$, each consisting of a total of 10 nodes. It's worth noting that in our article, we generate a set of 800 complex networks. $b_{1})$ Calculation of topological measures: In this step, we compute various topological measures for all the previously generated complex networks. However, for the sake of simplification in this illustration, we focus on a single measure: Betweenness Centrality (BC). We apply this calculation to one of the four networks, specifically network $i=4$. $b_{2})$ Implementation of the $Q$-voter model: In this stage, we implement the $Q$-voter model on each of the complex networks using three distinct initialization methods represented by colored circles: high-degree (purple), low-degree (green), and random selection (orange). This analysis is performed for both $Y_i$ (consensus time) and $C_i$ (frequency of opinion changes). For the sake of simplification, we select only network $i=4$ to illustrate this process. $c)$ Creation of the dataset: In this step, we construct a dataset that contains information from all the generated networks. Each row of the table represents a specific network, and the columns contain the calculations of topological measures for these complex networks. Additionally, we include the corresponding values for initialization methods (high-degree, low-degree, and random selection) regarding $Y_i$ and $C_i$. For illustration purposes, we present information only for network $i=4$, including BC and $Y_i$. However, in the full article, our table encompass 800 rows and 15 columns, comprising nine topological measures, along with three variations of initialization methods for $Y_i$ and $C_i$. $d)$ Application of machine learning algorithms: Finally, based on the gathered information, we apply machine learning algorithms to conduct further analyses and obtain significant insights and summary statistics from the data generated in the previous steps.}
\label{general_illustration}
\end{figure}

The schematic in Figure \ref{general_illustration} provides an overview of the comprehensive process outlined in this article, which encompasses several steps: $a)$ Generation of complex networks: We generate the eight types of networks under study. $b_{1})$ Calculation of topological measures: In this step, we compute the nine topological measures for all the previously generated complex networks. $b_{2})$ Implementation of the $Q$-voter model: In this stage, we implement the $Q$-voter model on each of the complex networks using three distinct initialization methods represented by colored circles: high-degree (purple), low-degree (green), and random selection (orange). This analysis is performed for both $Y_i$ (consensus time) and $C_i$ (frequency of opinion changes). $c)$ Creation of the dataset: A dataset is constructed containing information from all generated networks. Each row represents a specific network, and the columns contain topological measure calculations. The dataset also includes values for initialization methods (high-degree, low-degree, and random selection) for both $Y_i$ (consensus time) and $C_i$ (frequency of opinion changes). 
$d)$ Application of machine learning algorithms: Based on the collected information, machine learning algorithms are used to conduct further analyses and extract significant insights and summary statistics from the generated data.

\section{Results}
\label{results}

\begin{figure}[htbp]
\centering
\includegraphics[width=0.90\textwidth]{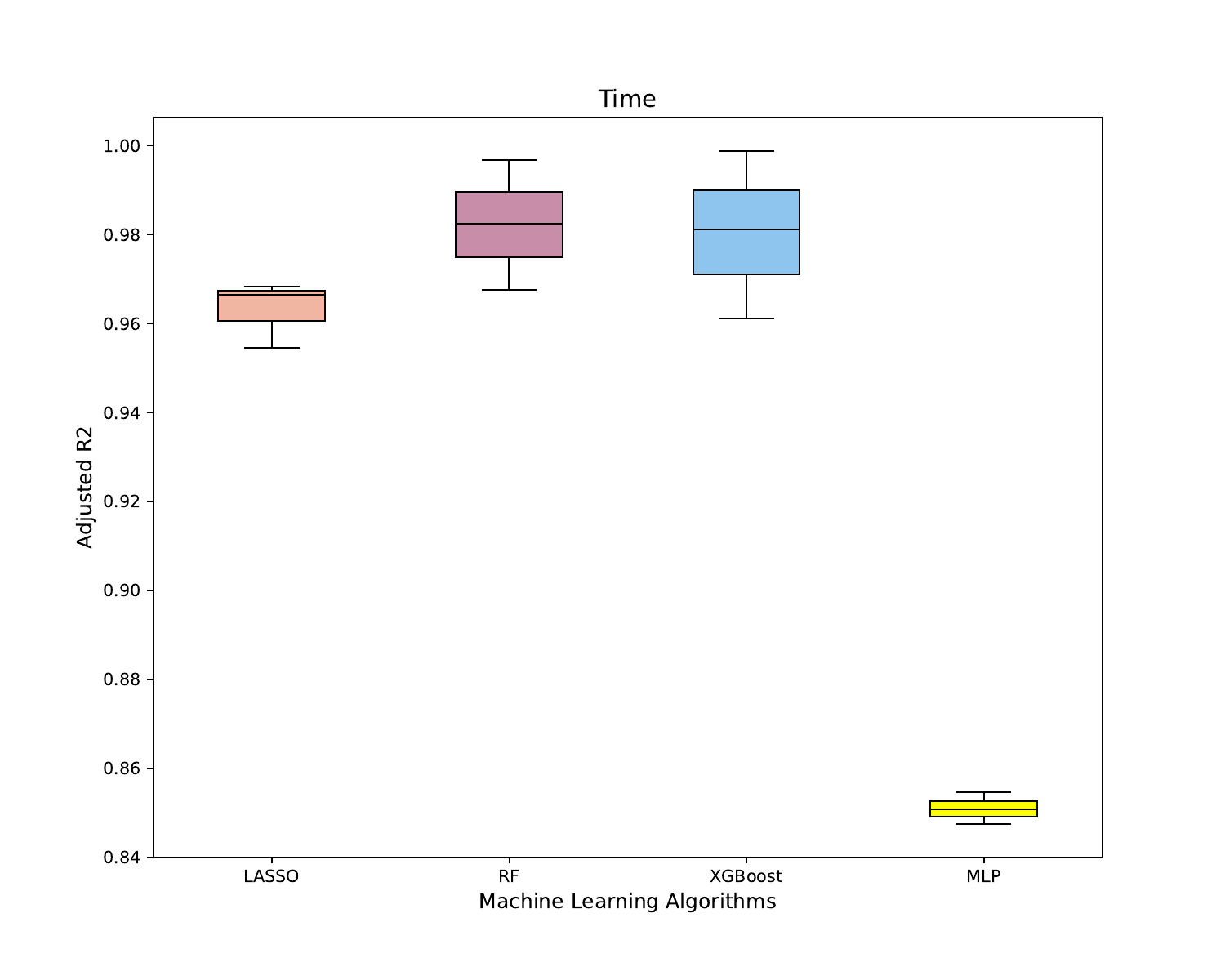}
\caption{Each boxplot represents the distribution of adjusted R2 values for the corresponding machine learning algorithms (LASSO, RF, XGBoost, and MLP), considering different initialization methods (high degree, low degree, and random selection) to predict $Y_{i}$. Among the algorithms, box 2 and box 3 correspond to the RF and XGBoost algorithms, respectively, and show the highest adjusted R2 values. This indicates that, on average, the RF and XGBoost algorithms outperform the other algorithms (LASSO and MLP) in terms of predictive accuracy.}
\label{boxplot_time}
\end{figure}

\begin{figure}[htbp]
\centering
\includegraphics[width=0.90\textwidth]{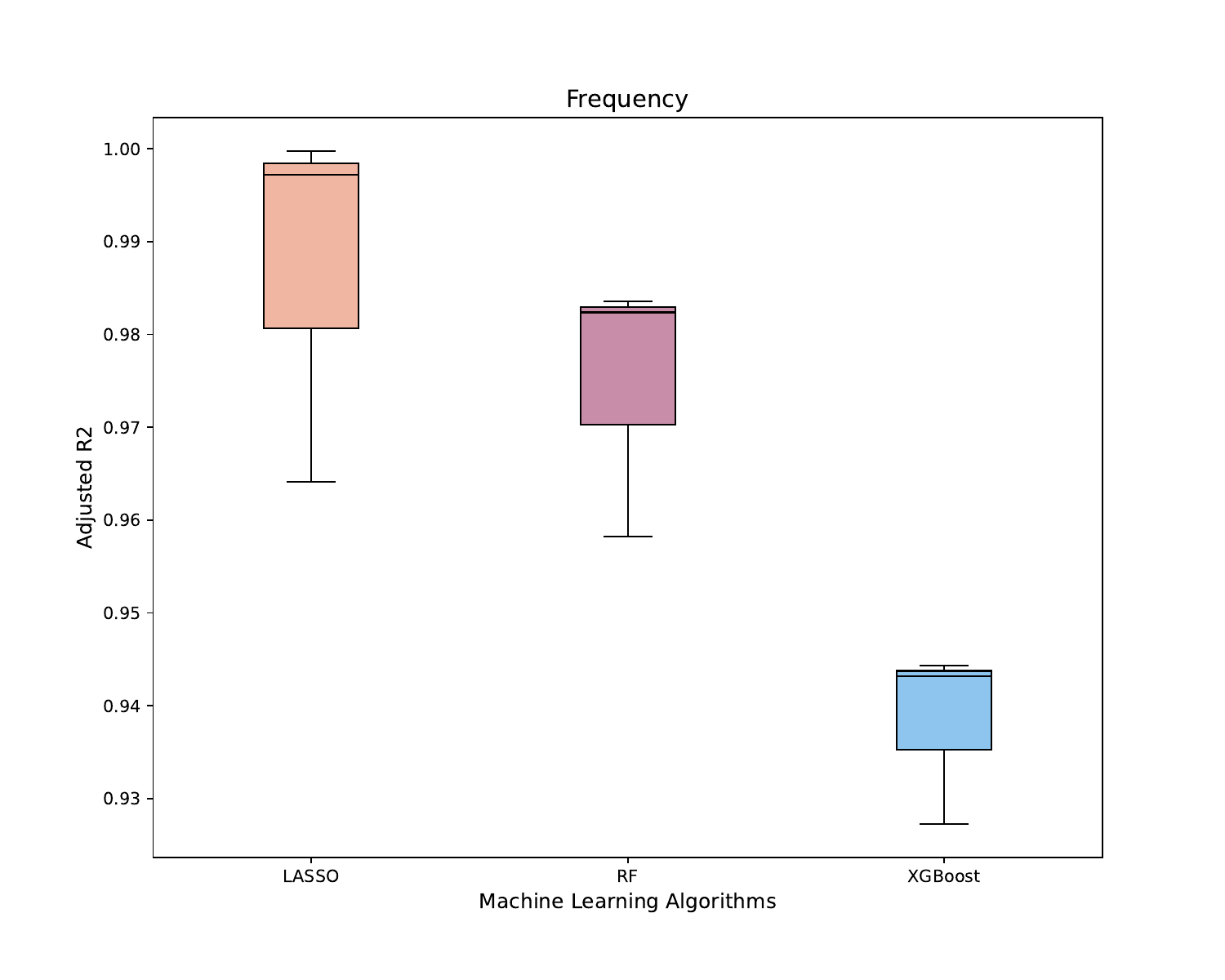}
\caption{Each boxplot represents the distribution of adjusted R2 values for the corresponding machine learning algorithm (LASSO, RF, and XGBoost), considering different initialization methods (high degree, low degree, and random selection) to predict $C_{i}$. Box 1, which corresponds to the LASSO algorithm, is the highest. This indicates that, on average, the adjusted R2 values for the LASSO algorithm are higher compared to the other algorithms (RF and XGBoost) considered.}
\label{boxplot_frequency}
\end{figure}

Figure \ref{boxplot_time} presents boxplots illustrating four machine learning algorithms: LASSO (light brown box), RF (pink box), XGBoost (blue box), and MLP (yellow box) for predicting $Y_{i}$. It is worth noting that RF (box 2, pink) and XGBoost (box 3, blue) exhibit the tallest boxes, indicating their tendency to yield higher average adjusted R2 values compared to the other algorithms. Furthermore, LASSO, RF, and XGBoost consistently produce the best results across all initialization methods, including high degree, low degree, and random selection. These three algorithms were selected for further analysis to predict $C_{i}$, and the results are presented in Figure \ref{boxplot_frequency}. Remarkably, LASSO (box 1, light brown) and RF (box 2, pink) emerge as the tallest boxes, suggesting their inclination to yield higher average adjusted R2 values compared to XGBoost. For this reason, we chose the RF algorithm, which stood out as the best in both figures, to illustrate the following figures (Figure \ref{prediction} and Figure \ref{importance}).

In Figures \ref{prediction}-A and B, we refer to the variables $Y_i$ and $C_i$, respectively, and illustrate the relationship between predicted values (\^y) on the y-axis and their corresponding original values (y) on the x-axis. Each point in the plot represents a specific data instance, where the x-coordinate indicates the actual value, and the y-coordinate represents the predicted value. The red dotted line represents a linear regression model, which provides an approximation of the overall trend in the data, aiding in the visualization of our model's predictive performance. For $Y_i$ (Figure \ref{prediction}-(A)), we calculated Pearson's correlation coefficients, resulting in values of 0.998 for high-degree initialization (purple dots), 0.991 for low-degree initialization (green dots), and 0.990 for random selection (orange dots). Additionally, we computed the adjusted R2 values, which were 0.996, 0.982, and 0.968, respectively, for the same initialization methods. For $C_i$ (Figure \ref{prediction}-(B)), we also calculated Pearson's correlation coefficients, yielding values of 0.999 for high-degree initialization, 0.991 for low-degree initialization, and 0.991 for random selection. The adjusted R2 values were 0.997, 0.983, and 0.945, respectively. These results underscore the correlations observed between the original and predicted values for both $Y_i$ and $C_i$, regardless of the initialization method used.

The RF algorithm assessed the input variables (network features) in our model. It evaluates the significance of variables by observing the improvement they provide to the model when incorporated into decision trees. The prioritization of network features based on their average importance across different initialization methods, as depicted in Figure \ref{importance}, provides valuable insights into their predictive capabilities. In this analysis, the features were ranked according to their average importance, considering three initialization methods: high degree (purple bar), low degree (green bar), and random selection (orange bar). Upon analyzing the bar chart (Figure \ref{importance}), it becomes apparent that network features with higher average importance occupy the top positions. Notably, when attempting to predict $Y_{i}$, the clustering coefficient (C) emerges as the most significant measure (Figure \ref{importance}-A). This indicates that the network's structure, particularly the formation of cohesive groups, plays a crucial role in the speed of consensus attainment within the $Q$-voter model. In terms of $C_{i}$, information centrality (IC) stands out as the most relevant network measure (Figure \ref{importance}-B). This suggests that the dissemination and influence of information within the network play a fundamental role in the dynamics of opinion changes. These network measures play vital roles in predicting different aspects of the $Q$-voter model. These inferences underscore the significance of different network aspects concerning the various phenomena under study. While the C focuses on consensus formation, IC pertains to opinion changes. These findings offer valuable insights for comprehending and forecasting the behavior of voter models in broader contexts. In contrast, measures such as eigenvector centrality (EC), Degree Pearson correlation coefficient (PC), and subgraph centrality (SC) do not exhibit significant predictive capabilities in these scenarios. 

Also, note that, individually, the CLC (represented by the purple bar in Figure \ref{importance}-A) becomes more relevant in networks initialized with a high degree of connectivity, while the AC (indicated by the orange bar in Figure \ref{importance}-A) is more significant in the randomly initialized networks. CLC gains importance when the dynamics of the $Q$-voter model are initiated by selecting nodes with a higher degree, as it measures how easily a node can communicate or influence other nodes in the network. When starting the dynamics with high-degree nodes, these high-degree nodes can have a substantial influence on the spread of opinions, and CLC can capture this capacity for influence. Similarly, the significance of the AC centrality measure when initiating the dynamics of the $Q$-voter model by selecting nodes randomly may be related to the definition of this centrality measure and the dynamics of opinion propagation in the $Q$-voter model on a network. AC is a measure that reflects the efficiency with which a node can transmit information or influence others in the network. When the dynamics of the $Q$-voter model are initiated randomly, there is no initial preference for high-degree or low-degree nodes. Therefore, it is crucial to identify nodes that can effectively facilitate the spread of opinions throughout the network, and AC can highlight nodes playing an important role in this regard.

Finally, the learning curve was calculated specifically for the two best results achieved using the high-degree initialization method for $Y_i$ (adjusted R2 = 0.996) and $C_i$ (adjusted R2 = 0.997). By manipulating the size of the training set, the learning curve offers valuable insights into the model's predictive capabilities \cite{spadon2019reconstructing}. This approach provides the advantage of understanding how the model's performance improves as more training instances are used, focusing on the most promising initialization methods. The findings depicted in Figures \ref{learning_curve} indicate that the complete database is not indispensable for achieving the highest level of validation accuracy. Surprisingly, even with a mere 200 training instances, the model demonstrated exceptional performance. These results emphasize that a relatively smaller training set can still yield satisfactory results.

\begin{figure}
\centering
\begin{minipage}[b]{0.65\textwidth}
\centering
\includegraphics[width=\textwidth]{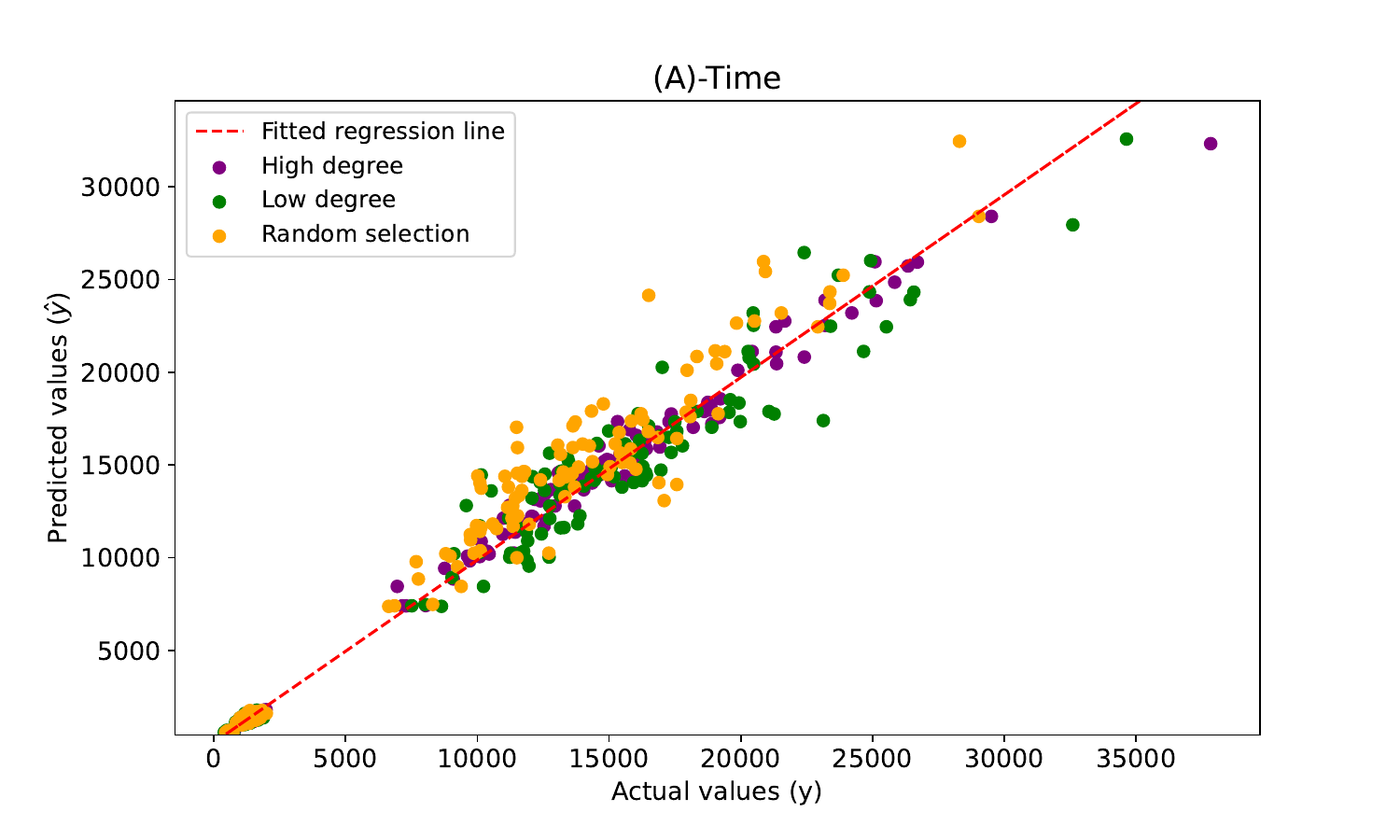}
\end{minipage}
\hfill
\begin{minipage}[b]{0.65\textwidth}
\centering
\includegraphics[width=\textwidth]{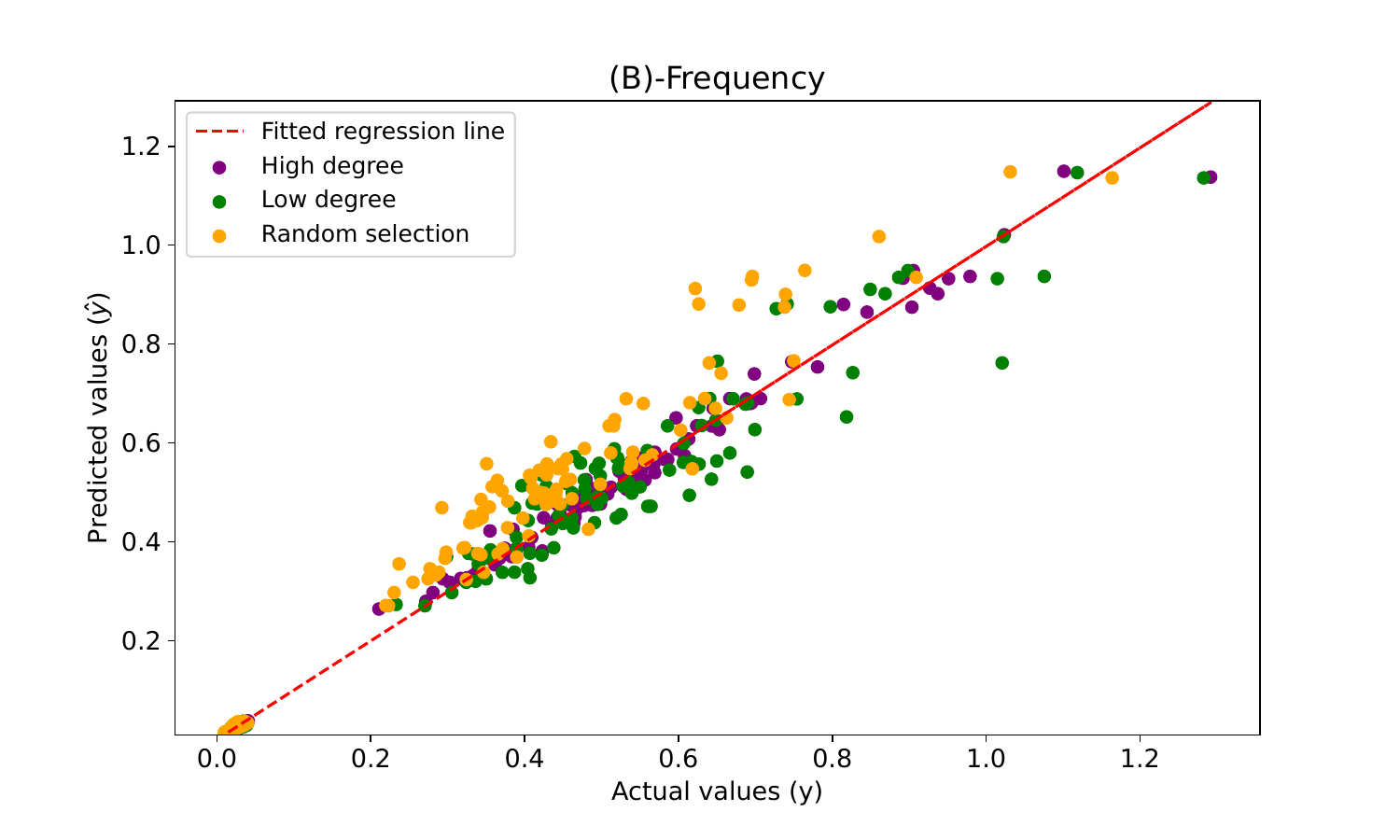}
\end{minipage}
\caption{Illustration showing the relationship between their corresponding original values (y) and predicted values (\^y) for (A) Time and (B) Frequency regarding the selection of agents with high degree (purple dots), low degree (green dots), and random selection (orange dots) for the initiation of dynamics. This analysis was conducted using the RF algorithm. This analysis was conducted using the RF algorithm.}
\label{prediction}
\end{figure}

\begin{figure}
\centering
\begin{minipage}[b]{0.45\textwidth}
\centering
\includegraphics[width=\textwidth]{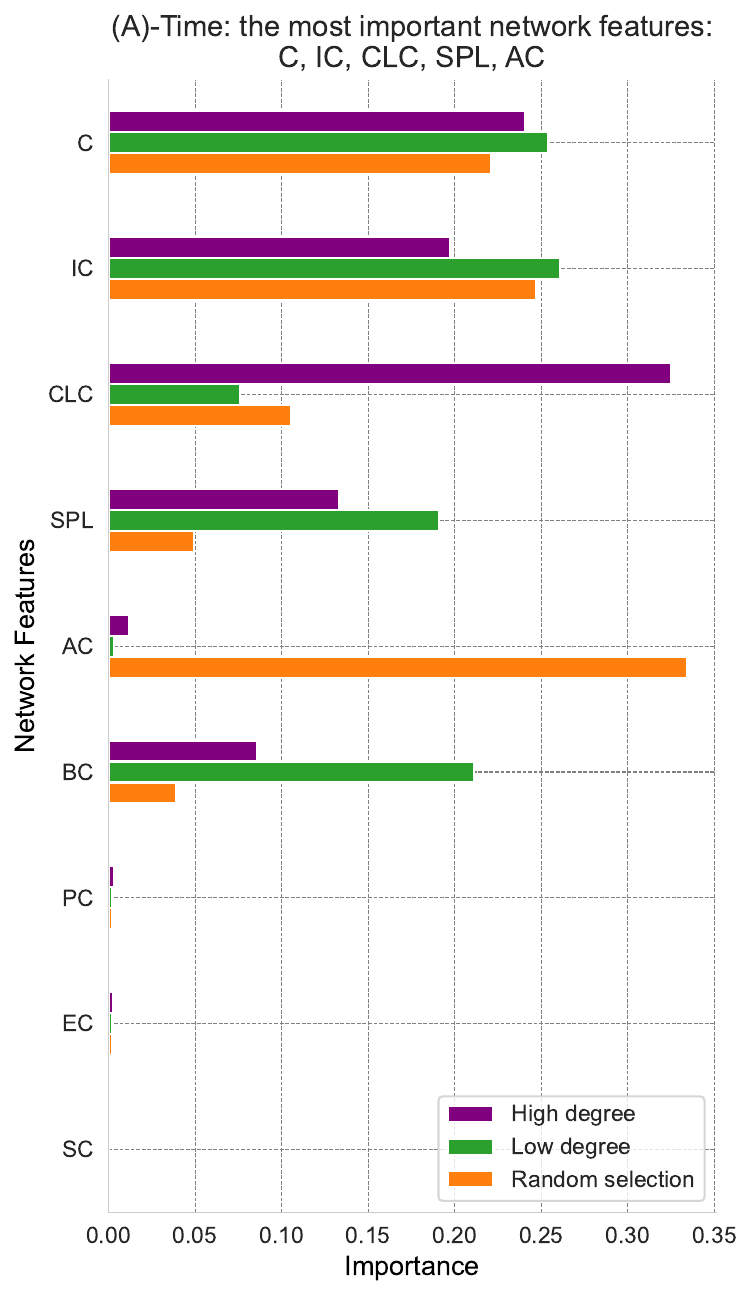}
\end{minipage}
\hfill
\begin{minipage}[b]{0.45\textwidth}
\centering
\includegraphics[width=\textwidth]{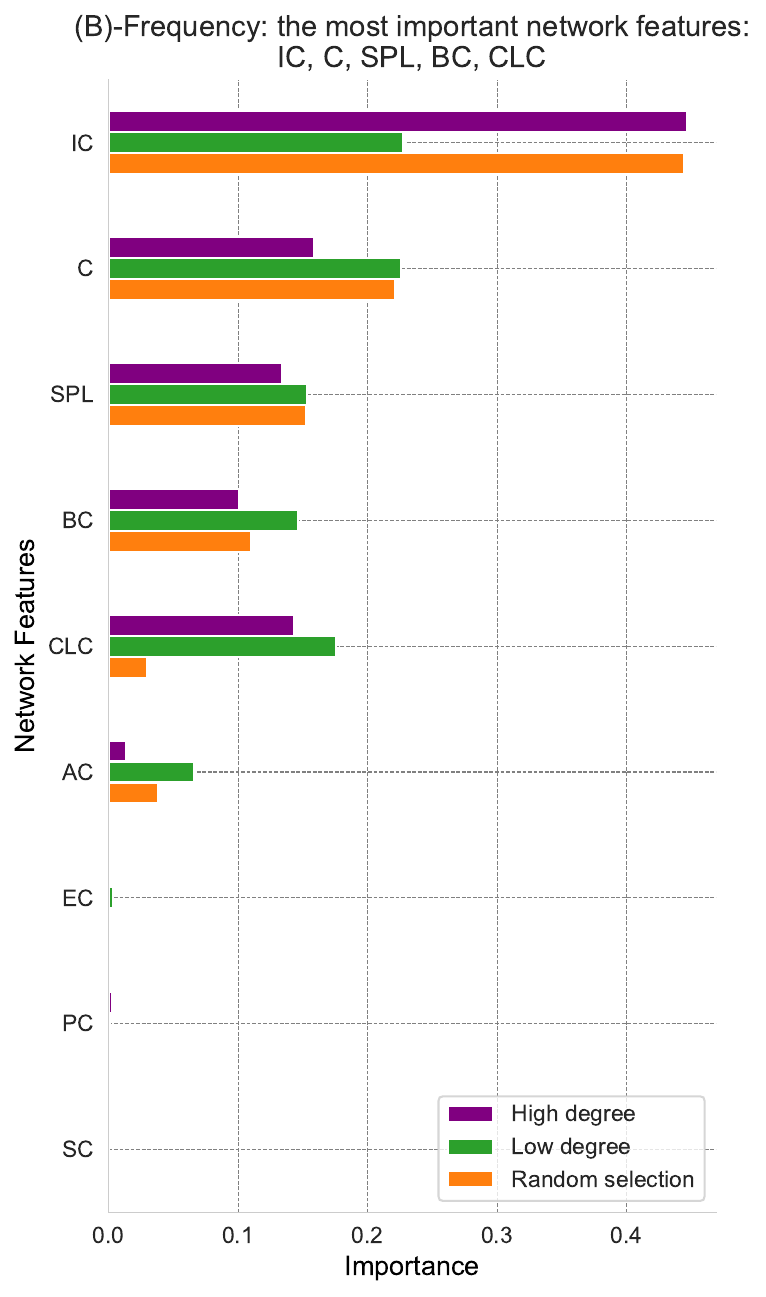}
\end{minipage}
\caption{The examination of the most crucial features, which are determined based on the average importance of complex network measures, was conducted to predict both (A) $Y_i$ and (B) $C_i$ using various initialization methods. These methods encompassed the selection of agents with the highest degree (purple bars), lowest degree (green bars), and random selection (orange bars) to initiate the dynamics. Notably, the clustering coefficient (C) and information centrality (IC) consistently emerged as the two most significant measures in both scenarios. This analysis was carried out employing the RF algorithm.}
\label{importance}
\end{figure}

\section{Conclusions}
\label{conclusions}

In this article, we predicted dynamic variables associated with $Q$-voter models based on network properties. We verified that the prediction is very accurate and determined which features most contribute to the emergence of polarization. Mainly, we show that the clustering coefficient and information centrality are the most important measures to quantify these patterns of connections. 
Moreover, variations in the initialization method, to start the dynamic of the $Q$-voter model with a positive opinion, were performed to predict consensus of the time ($Y_{i}$) and frequency of opinion changes ($C_{i}$). Initially, agents were randomly selected, following the original method of the $Q$-voter model. Subsequently, agents with the highest degree were identified and selected to investigate their potential for strongly influencing the overall opinion dynamics due to their extensive connections. Lastly, agents with the lowest degree of connectivity were considered initiators of the dynamics to explore the potential impact of less influential agents on opinion evolution. Although modifications in the initialization methods of positive opinions affect the results, their impact is relatively small. 
Indeed, subsequent interactions and information exchange among agents tend to overshadow the influence of the initially selected agents, leading to a consensus of opinions and a limited long-term impact of the initial agent selection. Nonetheless, the exploration of the role of both highly connected and less connected agents provided valuable insights into the complex dynamics of opinion formation and consensus emergence within the $Q$-voter model. 

We found that, regardless of the initialization method used to start the Q-voter model, the initial influence of the selected agents tends to decrease over time. This occurs because, as agents interact and exchange information, their opinions are influenced by others. Over time, opinions begin to converge towards a consensus, and the initial influence of randomly selected, high- or low-connectivity agents becomes equivalent since there is not a significantly superior initialization method over the others; all of them yield equally good results. When we say that the absence of influential agents contributes to a more efficient consensus, we mean that the absence of agents with disproportionate influence in the network means that each agent plays a similar role in shaping the collective opinion. This is important because polarization often occurs when a few extremely influential agents have a disproportionate impact on others' opinions. In the article \cite{centola2007complex}, the authors investigate the influence of highly connected individuals in opinion dynamics. Their research illustrates that a small number of highly connected individuals can significantly influence the polarization of opinions within a network. Furthermore, Sunstein's book `Republic: Divided Democracy in the Age of Social Media' \cite{sunstein2018republic} provides insights into the role of online platforms and highly influential users in shaping public discourse, potentially leading to polarization.

If all agents have similar influence, it is less likely that a few highly influential agents dominate the conversation and pull the collective opinion to opposite extremes. Therefore, the absence of highly influential agents can contribute to a more balanced and less polarized decision-making process.   

Expanding our methodology to explore the variance prediction within the $Q$-voter model can provide further insights into the factors that contribute to diverse outcomes in social dynamics. Future work in this direction will contribute to a more comprehensive understanding of the complex nature of polarization and its potential implications. By leveraging machine learning algorithms and complex network features, this study can advance research in the field of complex systems and pave the way for future investigations on the dynamics of polarization in various social contexts.
Overall, the combination of machine learning algorithms and complex network analysis has the potential to revolutionize our comprehension of social systems, leading to a deeper understanding of human behavior and the development of strategies that promote positive societal outcomes.

\section*{Acknowledgments}

A.M.P acknowledges the support of the São Paulo Research Foundation (FAPESP), grant 2021/13843-2. P. K. gratefully acknowledges support from the Engineering and Physical Sciences Research Council and Medical Research Council through the Mathematics of Systems I Centre for Doctoral Training at the University of Warwick (reference EP/L015374/1). F.A.R. acknowledges CNPq (grant 309266/2019-0) and FAPESP (grant 19/23293-0) for the financial support given for this research. This research was conducted with the computational resources of the  Center for Research in Mathematical Sciences Applied to Industry (CeMEAI) funded by FAPESP, grant 2013/07375-0.

\newpage
\appendix

\section{Construction of Complex Networks}
\label{Construction of Complex Networks}

In this appendix, parameters involved in network generation are presented in tabular format. The network values were adjusted to ensure that the average degree of all networks fell within the range of 9 to 10.

\begin{itemize}
\item \textbf{Erdős-Rényi:} We used the \texttt{nx.erdos\_renyi\_graph} function from NetworkX to create an Erdős-Rényi network \cite{networkxerdosrenyidoc}. The following table provides information concerning the creation of this network.

\begin{table}[ht]
\centering
\begin{tabular}{|l|l|}
\hline
\textbf{Parameter} & \textbf{Value} \\
\hline
$n$ & 1000 \\
$p$ & 0.01 \\
$seed$ & None \\
$directed$ & False \\
\hline
\end{tabular}
\caption{Parameters for the Erdős-Rényi network model.}
\label{tab:parameters}
\end{table}

\textbf{Parameter Descriptions:}

\begin{itemize}

\item \textbf{n}: The number of nodes in the network.

\item \textbf{p}: The probability for edge creation. The model chooses each of the possible edges with probability $p$.

\item \textbf{seed}: Indicator of random number generation state. In our case, it is set to None, which means the default random number generation state is used.

\item \textbf{directed}: If True, this function returns a directed network. In our case, it is set to False, indicating that the network is undirected.

\end{itemize}
\end{itemize}

\begin{itemize}
\item \textbf{Barabási Linear, Barabási Non-Linear (0.5), Barabási Non-Linear (1.5):} We employed the \texttt{graph.Barabasi} function to create networks following the Barabási-Albert model \cite{igraphbarabasidoc}. The subsequent table furnishes specific details regarding the generation of this network.

\begin{table}[ht]
\centering
\begin{tabular}{|l|l|}
\hline
\textbf{Parameter} & \textbf{Value} \\
\hline
$n$ & 1000 \\
$m$ & 5 \\
$outpref$ & True \\
$directed$ & False \\
$power$ & 1.0 \\
$zero\_appeal$ & 1 \\
$implementation$ & psumtree \\
$start\_from$ & None \\
\hline
\end{tabular}
\caption{Parameters for the Barabási-Albert Model.}
\label{tab:parameters}
\end{table}

\textbf{Parameter Descriptions:}

\begin{itemize}

\item \textbf{n}: The number of nodes in the generated network. In the example, 1000 nodes were created.

\item \textbf{m}: The number of outgoing edges generated for each node or a list containing the number of outgoing edges for each node explicitly. In the example, each node has 5 outgoing edges.

\item \textbf{outpref}: A boolean value that determines whether the out-degree of a node affects its citation probability. In the example, it is set to True.

\item \textbf{directed}: A boolean value that determines whether the generated network is directed. In the example, it is set to False, indicating that the network is undirected.

\item \textbf{power}: The power constant of the nonlinear model. In the example, the value is 1.0, representing the linear model.

\item \textbf{zero\_appeal}: The attractiveness of nodes with degree zero. In the example, it is set to 1.

\item \textbf{implementation}: The algorithm used to generate the network. In the example, it is set to psumtree, which uses a partial prefix-sum tree.

\item \textbf{start\_from}: If provided and not None, this parameter uses another network as a starting point for the preferential attachment model. In the example, no starting network is specified (None).

\end{itemize}
\end{itemize}

Note that to generate the Barabási networks in a non-linear manner, we modified the power parameter to 0.5 and later to 1.5.

\begin{itemize}
\item \textbf{LFR (Lancichinetti-Fortunato-Radicchi Benchmark)}: We generated LFR networks using the \texttt{LFR\_benchmark\_graph}  function \cite{networkxdoc}. A table following this one provides information on how this network was generated.

\begin{table}[ht]
\centering
\begin{tabular}{|l|l|}
\hline
\textbf{Parameter} & \textbf{Value} \\
\hline
$n$ & 1000 \\
$\tau_1$ & 3 \\
$\tau_2$ & 1.5 \\
$\mu$ & 0.1 \\
$average\_degree$ & 10 \\
$min\_degree$ & None \\
$max\_degree$ & None \\
$min\_community$ & 100 \\
$max\_community$ & None \\
$tol$ & $1 \times 10^{-7}$ \\
$max\_iters$ & 500 \\
$seed$ & 10 \\
\hline
\end{tabular}
\caption{Parameters for the LFR Benchmark network model.}
\label{}
\end{table}

\textbf{Parameter Descriptions:}
\begin{itemize}

\item\textbf{n}: Number of nodes in the created network.

\item\textbf{$\tau_1$}: Power law exponent for the degree distribution of the created network. This value must be strictly greater than one.

\item\textbf{$\tau_2$}: Power law exponent for the community size distribution in the created network. This value must be strictly greater than one.

\item\textbf{$\mu$}: Fraction of inter-community edges incident to each node. This value must be in the interval $[0, 1]$.

\item\textbf{average\_degree}: Desired average degree of nodes in the created network. This value must be in the interval $[0, n]$.

\item\textbf{min\_degree}: Minimum degree of nodes in the created graph. This value must be in the interval [0, n].

\item\textbf{max\_degree}: Maximum degree of nodes in the created network. If not specified, this is set to $n$, the total number of nodes in the network.

\item\textbf{min\_community}: Minimum size of communities in the network. If not specified, this is set to min\_degree.

\item\textbf{max\_community}: Maximum size of communities in the network. If not specified, this is set to $n$, the total number of nodes in the network.

\item\textbf{tol}: Tolerance when comparing floats, specifically when comparing average degree values.

\item\textbf{max\_iters (int)}: The maximum number of iterations to attempt in order to create community sizes, degree distribution, and community affiliations.

\item\textbf{seed(integer, random\_state, or None - default)}: 
An indicator of the random number generation state.

\end{itemize}
\end{itemize}

\begin{itemize}

\item \textbf{Watts-Strogatz:} We used the \texttt{nx.watts\_strogatz\_graph} from the NetworkX library to generate a Watts-Strogatz network \cite{networkxwattsstrogatzdoc}. The following table contains information about the values of each parameter of this network.

\begin{table}[h]
\centering
\begin{tabular}{|c|c|}
\hline
\textbf{Parameter} & \textbf{Value} \\
\hline
n &  1000 \\
k & 10 \\
p & 0.01 \\
\hline
\end{tabular}
\caption{Parameters for the Watts-Strogatz network model.}
\end{table}

\textbf{Parameter Descriptions:}
\begin{itemize}
    \item \textbf{n}: The number of nodes.
      \item \textbf{k}: Each node is joined with its k nearest neighbors in a ring topology.
      \item \textbf{p}: The probability of rewiring each edge.
\end{itemize}
\end{itemize}

\begin{itemize}
    \item \textbf{Waxman:} We used the \texttt{nx.waxman\_graph} function from the NetworkX library to generate a Waxman network \cite{networkxwaxmandoc}.

\begin{table}[h]
\centering
\begin{tabular}{|c|c|}
\hline
\textbf{Parameter} & \textbf{Value} \\
\hline
n & 1000 \\
beta & 0.12 \\
alpha & 0.1 \\
L & None \\
domain & (0, 0, 1, 1) \\
metric & function \\
seed & None (default) \\
\hline
\end{tabular}
\caption{Parameters for the Waxman network model.}
\end{table}

\textbf{Parameter Descriptions:}

 \begin{itemize}
\item \textbf{n}: Number of nodes.
\item \textbf{$beta$}: Model parameter.
\item \textbf{$alpha$}: Model parameter.
\item \textbf{L}: The maximum distance between nodes is set to be the maximum distance between any pair of nodes.
\item \textbf{domain}: Domain size, given as a tuple of the form ($x\_min$, $y\_min$, $x\_max$, $y\_max$).
\item \textbf{metric}: Euclidean distance metric is used. 
\item \textbf{seed (integer, random\_state, or None)}: Indicator of random number generation state (default is None).

\end{itemize}
\end{itemize}

\begin{itemize}
    \item \textbf{Path:} We used the \texttt{nx.path\_graph} function from the NetworkX library to generate this network \cite{networkxpathgraphdoc}. Note that in our code available on GitHub \href{https://github.com/kentwar/QVML_2023/}{here} for generating networks, we have added specific lines of code for the path graph to ensure that the average degree falls within the range of 9 to 10, aligning with the characteristics of the other networks generated.

\begin{table}[h]
\centering
\begin{tabular}{|c|c|}
\hline
\textbf{Parameter} & \textbf{Value} \\
\hline
n & 1000 \\
\hline
\end{tabular}
\caption{Parameters for the Path network model.}
\end{table}

\textbf{Parameter Descriptions:} 
\begin{itemize}
    \item\textbf{n}: Number of nodes.
\end{itemize}
\end{itemize}

\section{Network Measurement Details}
\label{Network Measurement Details}

\subsection{\textbf{Clustering coefficient (C)}}

The local clustering coefficient (C) is an important metric in network and graph analysis that quantifies the tendency of neighbors of a node in a network to cluster together. In other words, it measures the degree of connectivity among the direct neighbors of a specific node, which is useful for understanding community structure and cohesion within a network. The mathematical formula for calculating C of a node $v$ in a graph is as follows:

\begin{equation}
    C(v)= \frac{2*E(v)}{k_{v}*(k_{v} - 1)}
\end{equation}

where:
\begin{itemize}
    \item $C(v)$ is the local clustering coefficient of node $v$.
    \item $E(v)$ is the number of edges between the direct neighbors of $v$ (i.e., the triangles that include node $v$).
    \item $k_{v}$ is the degree of node $v$, which is the number of direct neighbors it has.
\end{itemize}

The \textbf{transitivity\_local\_undirected(mode="zero")} is a Python function commonly employed in network analysis using the Igraph library. This function calculates the C for individual nodes within a graph. It operates in "zero" mode, which specifically considers triangles in the network that share exactly one node with the node being analyzed. The output of this function is a data structure, typically a list or a similar container, containing the C corresponding to each node in the graph. Finally, we calculate the mean to get a final value. 

\subsection{\textbf{Closeness Centrality (CLC)}}

Local closeness centrality (CLC) is a network analysis metric that measures how close a node is to all the other nodes in its local neighborhood within a graph. It quantifies how quickly information can spread from a specific node to its neighboring nodes. Nodes with higher CLC are considered to be more central within their local environment, as they can reach other nodes more efficiently. The mathematical formula for the CLC of a node $v$ is as follows:

\begin{equation*}
    CLC(v)= \frac{1}{\sum_{u \neq v}} d(v,u)
\end{equation*}

where:
\begin{itemize}
    \item $CLC(v)$ is the local closeness centrality of node $v$. 
    \item $d(v,u)$ represents the shortest path distance between nodes $v$ and $u$ in the graph. The $\sum$ in the denominator calculates the sum of the shortest path distances from node $v$ to all other nodes $u$ in its local neighborhood.
\end{itemize}

The \textbf{closeness\_centrality(normalized=True)} function is a commonly used Python function in network analysis using the Igraph library. This function calculates CLC measures for each node in a graph. When we use normalized=True, it indicates that we want the CLC values to be normalized. In other words, the values are adjusted to be within the range of 0 to 1, making these measures comparable across different graphs, regardless of the network's size or scale. Finally, by calculating the average of these normalized measures, we obtain a representative value of the average closeness centrality in the network, which is useful for assessing the communication efficiency of nodes within their respective local environments.

\subsection{\textbf{Betweenness Centrality (BC)}}

Betweenness Centrality (BC) is a fundamental metric in network analysis that assesses the importance of nodes as crucial intermediaries in communications within a network. Mathematically, the formula for calculating the BC of a node is as follows:

\begin{equation*}
    BC(v)= \sum_{s \neq v \neq t} \frac{\phi_{st}(v)}{\phi_{st}}
\end{equation*}

where:
\begin{itemize}
    \item $BC(v)$ is the betweenness centrality of node $v$. 
    \item $\phi_{st}$ is the total number of shortest paths (geodesics) between nodes $s$ and $t$.
    \item $\phi_{st}(v)$ is the number of shortest paths between $s$ and $t$ that pass through node $v$.
\end{itemize}

The \textbf{betweenness\_centrality()} function is a specific feature of the NetworkX library, widely used for network analysis in Python. This function is responsible for calculating BC in a graph. Essentially, it assesses the importance of each node within the graph by measuring how often a node acts as a crucial bridge in the shortest paths between other nodes in the network. The result of this function is a dictionary where the keys represent the nodes in the graph, and the corresponding values are the BC measures associated with these nodes. This analysis is valuable for identifying nodes that play a critical role as intermediaries in communication or the transportation of information within a network.

\subsection{\textbf{Shortest path length (SPL)
}}
The Shortest Path Length (SPL), also known as the length of the shortest path, is a metric that describes the distance between two nodes in a graph, representing the minimum number of edges or weighted edges required to travel from node A to node B within the network. The formula to calculate the SPL between two nodes can be described as: 

\begin{itemize}
    \item  SPL(A, B) = the smallest number of edges between nodes A and B.
\end{itemize}
  
In Python, we can calculate the SPL libraries such as Igraph. For example, the \textbf{average\_path\_length()} function in calculates the average shortest path length between nodes in the network, providing a valuable measure for evaluating the efficiency of transportation, communication, and connectivity in a network.

\subsection{\textbf{Degree Pearson correlation coefficient (PC)
}}
The Pearson Correlation Coefficient for Degrees (PC) is a metric that assesses the linear relationship between the degrees of nodes in a graph. It measures the tendency of nodes with similar degrees to connect or whether they prefer to link to nodes with different degrees. This measure is important for understanding how the network is organized in terms of node degrees, indicating whether there is a tendency for assortativity (positive correlation) or disassortativity (negative correlation) in the network's connectivity. The formula for calculating the PC is given by:

\begin{equation*}
    PC= \frac{\sum_{(x_i - \hat{x})*(y_i- \hat{y})} }{\sum(x_i - \hat{x})^2 * \sum (y_i - \hat{y})^2} 
\end{equation*}

where:

\begin{itemize}
    \item $PC(v)$ is the Pearson Correlation Coefficient. 
    \item $x_i$ and $y_i$ are the degrees of the nodes. 
    \item $\hat{x}$ and $\hat{y}$ are the means of the node degrees.
\end{itemize}

In Python, we can calculate the Pearson Correlation Coefficient for Degrees using libraries such as NetworkX. The functions \textbf{degree\_pearson\_correlation\_coefficient()} in NetworkX can be used to calculate this measure on a graph represented by the respective libraries. The result will inform us about the nature of the network's connectivity about node degrees, which is useful for network analysis and characterization.

\subsection{\textbf{Information centrality (IC)
}}

Information centrality (IC) is a network metric used to assess the importance of nodes in a graph in terms of how they facilitate the flow of information or communication within the network. This metric is based on the idea that some nodes may act as critical points for the efficient dissemination of information in a network. Information centrality measures the amount of information a node is capable of controlling or transmitting to other nodes in the network. The mathematical formula for IC is defined as:

\begin{equation*}
    IC(v)= \sum_{u \neq v}\frac{1}{d(v,u)}
\end{equation*}

where:
\begin{itemize}
    \item $IC(v)$ is the information centrality of node $v$. 
    \item $\sum$ represents the sum over all nodes $u$ different from $v$. 
    \item $d(v,u)$ is the geodesic distance between nodes $u$ and $v$, i.e., the length of the shortest path between them.
\end{itemize}
This formula calculates the information centrality of a node by summing the inverses of the geodesic distances between the node in question $v$ and all other nodes $u$ in the graph. The shorter the path between $v$ and $u$, the greater the contribution of node 
$u$ to the information centrality of $v$. Therefore, nodes that are closer to $v$ will have a higher contribution to its information centrality.

In Python, we can use the \textbf{information\_centrality()} function from NetworkX to calculate the IC for the nodes in a graph. The function returns a dictionary where the keys are the nodes in the graph, and the values are the corresponding information centrality scores. This allows us to identify the most critical nodes in the network in terms of their ability to influence the flow of information.

\subsection{\textbf{Subgraph centrality (SC)
}}
Subgraph centrality is a network centrality (SC) metric that assesses the importance of a node based on how many subgraphs containing that node are connected in the network. In other words, it measures how central a node is in terms of its participation in interconnected subgraphs. The mathematical formula for SC is defined as follows:

\begin{equation}
SC(v) = \sum_{S \subseteq N \setminus \{v\}} \left(\frac{1}{1 + |E(S)|}\right)
\end{equation}

where:

\begin{itemize}
    \item $SC(v)$ is the subgraph centrality of node $v$.
    \item $S$ is a subset of the neighbors of $v$.
    \item $N$ is the set of neighbors of $v$.
    \item $E(S) $ is the number of edges in the subgraph induced by $S$.
\end{itemize}

This formula calculates the SC of a node $v$ by summing the contributions of all subsets of its neighbors. The more subsets contain $v$, and the more these subsets are interconnected (have fewer edges), the higher the subgraph centrality of $v$.
In Python, we can use the \textbf{subgraph\_centrality()} function from NetworkX to calculate the SC for the nodes in a graph. The function returns a dictionary where the keys are the nodes in the graph, and the values are the corresponding SC scores. This allows us to identify nodes that play a crucial role in connecting interconnected subgraphs in the network. Keep in mind that the calculation can be computationally expensive in large networks due to the need to evaluate many subsets of neighbors for each node.

\subsection{\textbf{Approx. Current flow betweenness centrality (AC)
}}

Approximate current flow betweenness centrality is a metric that assesses the importance of nodes based on their ability to influence the flow of electrical current within a network. Unlike the traditional approach to betweenness centrality, which precisely calculates exact paths, this methodology employs numerical methods, such as Monte Carlo algorithms, to estimate the flow of current between all pairs of nodes in the network. This approach makes it suitable for large-scale and complex networks. To calculate this centrality metric in Python, we use the \textbf{approximate\_current\_flow\_betweenness\_centrality} function from the NetworkX library. The result is a dictionary that associates each node in the network with its approximate centrality value. This metric plays a vital role in network analysis across various domains, aiding in the identification of key points of control and influence.

\subsection{\textbf{Eigenvector centrality (EC)
}}
Eigenvector centrality (EC) is a measure of centrality in a network or graph that assesses the relative importance of a node based on its connections to other nodes in the network. The underlying idea is that nodes connected to other important nodes are themselves important. Therefore, eigenvector centrality takes into account not only the number of connections a node has but also the importance of the nodes to which it is connected.
The mathematical formula to calculate the EC of a node in a graph is defined by the following equation:

\begin{equation*}
EC(v) = \frac{1}{\lambda} \sum_{u \in N(v)} w(u, v) \cdot C(u)
\end{equation*}

where:

\begin{itemize}
\item $EC(v)$ is the eigenvector centrality of node $v$.
\item $\lambda$ is the eigenvalue associated with the largest eigenvalue of the adjacency matrix of the graph.
\item $\sum$ represents the sum over all nodes $u$ connected to node $v$.
\item $w(u,v)$ is the weight of the edge between nodes $u$ and $v$.
\item $C(u)$ is the eigenvector centrality of node $u$.
\end{itemize}

The \textbf{eigenvector\_centrality()} function is part of the Igraph library in Python, used to calculate eigenvector centrality in a graph. EC is a measure that assesses the importance of nodes in a graph based on their connections, taking into account the importance of the nodes to which they are connected. The EC values are not scaled, meaning they reflect the raw measure of importance for each node in the graph. To obtain a single centrality measure for the entire graph, it's common to calculate the average of the centrality values for all nodes.

The Python code used to generate the $Q$-voter model, as well as the complex networks and measures of complex networks, is available for access at \cite{qvotergithub}.

\section{Principal Component Analysis (PCA)}
\label{apendix:PCA}

\begin{figure}
\centering
\begin{minipage}[b]{0.65\textwidth}
\centering
\includegraphics[width=\textwidth]{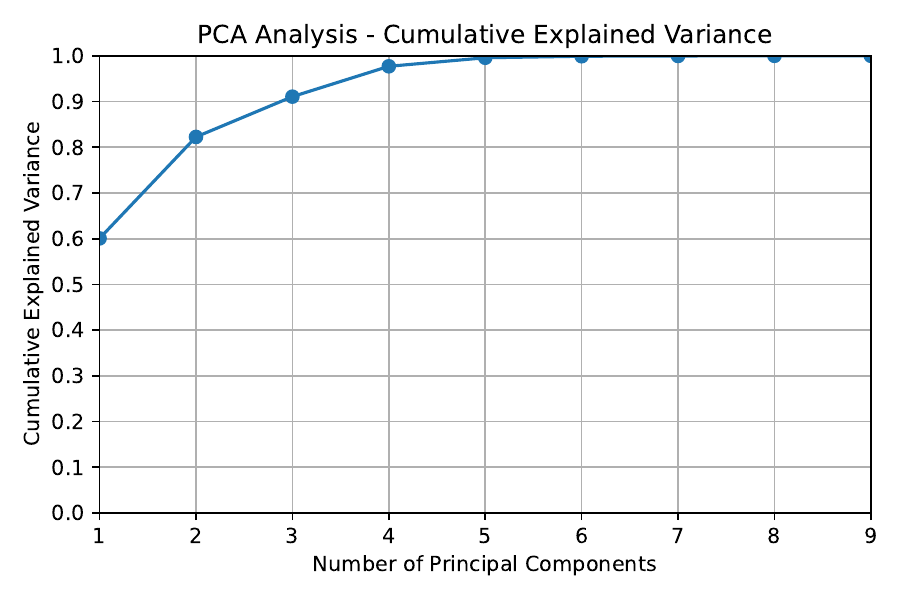}
\end{minipage}
\hfill
\begin{minipage}[b]{0.65\textwidth}
\centering
\includegraphics[width=\textwidth]{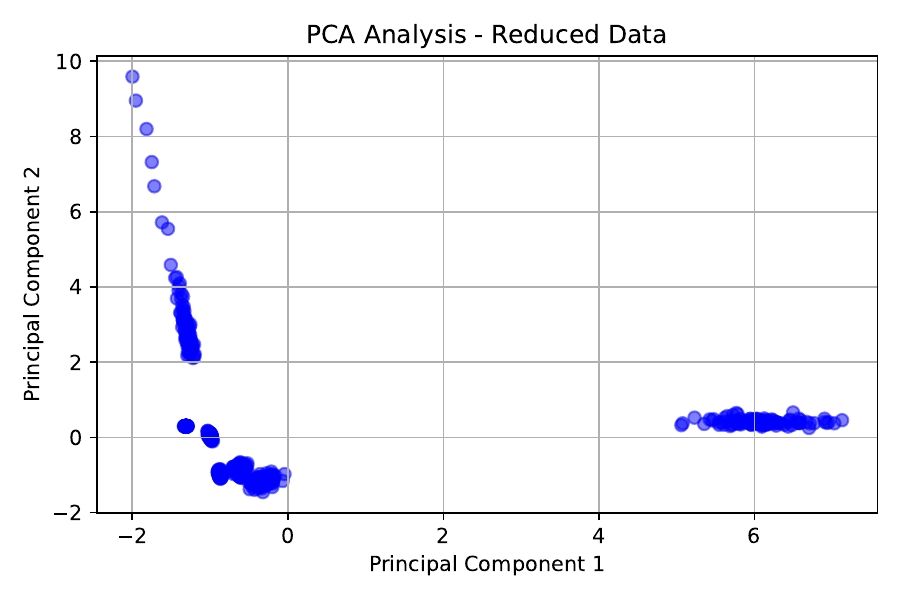}
\end{minipage}
\caption{The top figure illustrates the Cumulative Explained Variance in PCA (Principal Component Analysis) Analysis. This plot showcases the cumulative amount of variance in the data explained by each principal component, while the subsequent figure displays the Reduced Data Plot using Principal Components. The reduced data is represented in a lower-dimensional space defined by the principal components, allowing for a simplified representation of the original data while preserving its underlying structure. These figures provide insights into the data used to feed our machine-learning prediction models and demonstrate the effectiveness of PCA in reducing the dimensionality of the input data.}
\label{PCA}
\end{figure}

The analysis of cumulative explained variance provides valuable insights into the dimensionality reduction achieved by the PCA algorithm. The plot of cumulative explained variance illustrates the amount of information retained as the number of principal components increases (Figure \ref{PCA}). This information helps determine the minimum number of principal components required to capture a significant portion of the original data's variability, considering the dataset with 800 rows and 9 columns. This analysis is crucial for making decisions regarding the dimensionality reduction process,  in the context of changing network topologies every 100 rows.

On the other hand, the plot of the reduced data using the principal components visually represents the transformed dataset in a lower-dimensional space (Figure \ref{PCA}). By visualizing the data in this reduced space, which is particularly important in the case of high-dimensional data with 9 complex network measures, a better understanding of its structure and potential patterns or clusters that may exist is gained. These plots play a vital role in validating the effectiveness of the PCA algorithm in capturing the most relevant features of the data while reducing its dimensionality, considering the complexity and diversity of the network measures across different network topologies.

Additionally, the proximity of data points in the reduced space reflects the similarity between the models, allowing for the identification of clusters or groupings within each network topology and across different topologies. This further aids in understanding the relationships and similarities among different instances in the dataset, facilitating comparative analysis and identification of common characteristics or trends. Overall, these plots provide valuable insights into the data, aiding in analysis, interpretation, and model comparison, particularly in the context of complex networks with multiple measures and changing topologies.

\section{Grid search hyperparameter tuning}
\label{apendix:grid}

Table \ref{table-auti:gridsearch} shows the hyperparameters optimized by grid search.

\begin{table*}[htbp]
\caption{Hyperparameters for each machine learning algorithm optimized by grid search optimizer.}
\label{table-auti:gridsearch}
 \begin{adjustbox}{width=\textwidth}
\begin{tabular}{cll}
\hline
\textbf{Predictor} & \multicolumn{1}{c}{\textbf{Hyperparameters and description}}                                                           & \multicolumn{1}{c}{\textbf{Values}} \\ 
\hline
\textbf{RF}  & \begin{tabular}[c]{@{}l@{}}- max\_depth:  Maximum depth of the tree.\\ - max\_features: Number of features to be considered \\   toward a best split.
 \\ - min\_samples\_leaf : Minimum number of \\ samples required to be at a leaf node.\\ - min\_samples\_split: Minimum number of \\ samples for the split of an internal node.
\\ - n\_estimators: Number of trees in the forest.\end{tabular}                                      & 

\begin{tabular}[c]{@{}l@{}}{[}10,20,30,40,50{]}\\ {[}2,3,4{]}\\ \\ {[}1,2,4{]}\\ \\ {[}2,5,10{]}\\ \\ {[}100,200,300{]}\end{tabular}                                                                
\\ \hline

\textbf{LASSO}         & \begin{tabular}[c]{@{}l@{}}- regularization parameter.\end{tabular}                      & range 0.0001 to 0.0005                                                                                              \\ \hline

\textbf{MLP}        & \begin{tabular}[c]{@{}l@{}}- activation: Activation function for the hidden layer.\\ - solver: Solver for weight optimization.\\ - alpha:  L2 penalty (regularization term) parameter.\\ - batch\_size: Size of minibatches for stochastic optimizers.\\ - learning\_rate: Learning rate schedule for weight updates.\\ - learning\_rate\_init: Initial learning rate used.\end{tabular}                                    & \begin{tabular}[c]{@{}l@{}}{[}identity, logistic, tanh, relu{]}\\ {[}lbfgs, sgd, adam{]} \\ {[}0.0001,1e-5,0.01,0.001{]}\\ {[}1000,5000{]}\\ {[}constant, invscaling, adaptive{]} \\ {[}0.001,0.01,0.1,0.2,0.3{]}\end{tabular}

\\ \hline

\textbf{XGBoost}        & \begin{tabular}[c]{@{}l@{}}- subsample: fraction of observations to be \\ randomly sampled in each tree. \\ - max\_depth: maximum depth of each tree.\end{tabular}           
 & 
\begin{tabular}[c]{@{}l@{}}{[}0.6,0.8,1.0{]}\\\\{[}3,4,5{]}\end{tabular}               
                                         \\ 
                                         \hline
\end{tabular}
\end{adjustbox}
\end{table*}

\begin{figure}
\centering
\begin{minipage}[b]{0.65\textwidth}
\centering
\includegraphics[width=\textwidth]{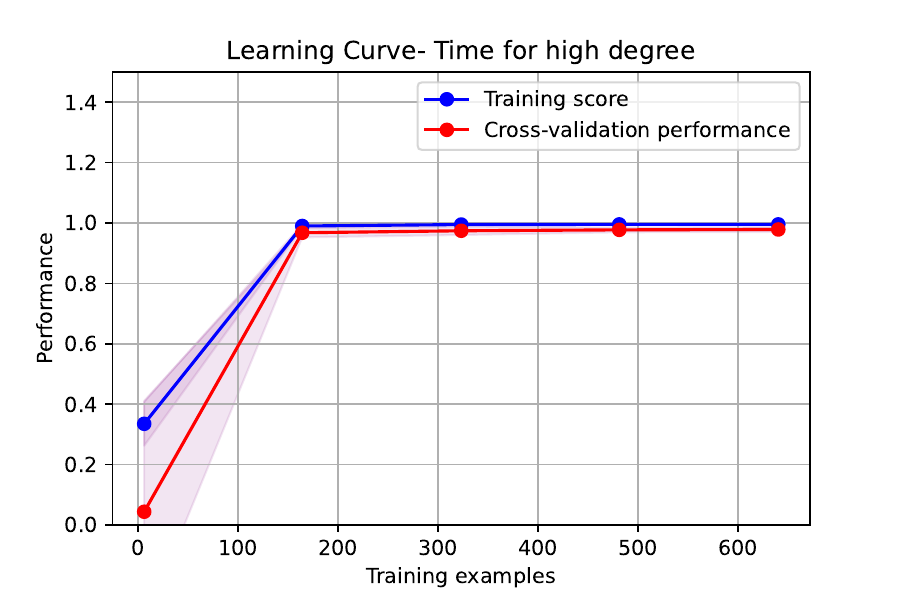}
\end{minipage}
\hfill
\begin{minipage}[b]{0.65\textwidth}
\centering
\includegraphics[width=\textwidth]{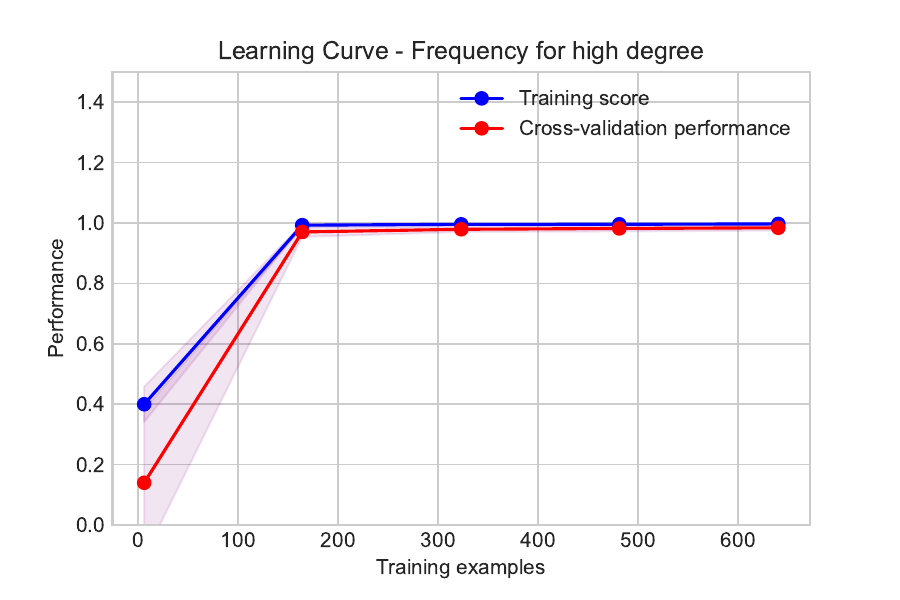}
\end{minipage}
\caption{The learning curve displays the training accuracy (represented by the blue curve) and the test accuracy (represented by the red curve) for the high initialization method. The top figure illustrates the learning curve for predicting $Y_{i}$, while the subsequent figure shows the learning curve for predicting $C_{i}$. These learning curves offer valuable insights into the performance of the model and the effectiveness of the high initialization method. By examining the training and test accuracies, one can evaluate the model's ability to generalize to unseen data and detect potential concerns such as overfitting or underfitting.}
\label{learning_curve}
\end{figure}

\newpage
\bibliographystyle{iopart-num} 
\bibliography{references}

\providecommand{\noopsort}[1]{}\providecommand{\singleletter}[1]{#1}%
\providecommand{\newblock}{}
\begin{thebibliography}{10}
\expandafter\ifx\csname url\endcsname\relax
  \def\url#1{{\tt #1}}\fi
\expandafter\ifx\csname urlprefix\endcsname\relax\def\urlprefix{URL }\fi
\providecommand{\eprint}[2][]{\url{#2}}

\bibitem{Thurner18}
Thurner S, Hanel R and Klimek P 2018 {\em Introduction to the theory of complex
  systems\/} (Oxford University Press)

\bibitem{Boccara10}
Boccara N and Boccara N 2010 {\em Modeling complex systems\/} vol~1 (Springer)

\bibitem{del2016spreading}
Del~Vicario M, Bessi A, Zollo F, Petroni F, Scala A, Caldarelli G, Stanley H~E
  and Quattrociocchi W 2016 {\em Proceedings of the National Academy of
  Sciences\/} {\bf 113} 554--559

\bibitem{flaxman2016filter}
Flaxman S, Goel S and Rao J~M 2016 {\em Public opinion quarterly\/} {\bf 80}
  298--320

\bibitem{barbera2015tweeting}
Barber{\'a} P, Jost J~T, Nagler J, Tucker J~A and Bonneau R 2015 {\em
  Psychological science\/} {\bf 26} 1531--1542

\bibitem{Conover2011}
Conover M~D, Ratkiewicz J, Francisco M, Goncalves B, Menczer F and Flammini A
  2011 {\em ICWSM\/} {\bf 133} 89--96

\bibitem{bessi2016social}
Bessi A and Ferrara E 2016 {\em Journal of Computer-Mediated Communication\/}
  {\bf 21} 303--320

\bibitem{centola2018experimental}
Centola D 2018 {\em Science\/} {\bf 360} 1116--1119

\bibitem{castellano2009statistical}
Castellano C, Fortunato S and Loreto V 2009 {\em Reviews of Modern Physics\/}
  {\bf 81} 591

\bibitem{soares2021empirical}
Soares G, Oliveira S and Carvalho T~C 2021 {\em arXiv preprint
  arXiv:2103.14179\/}

\bibitem{soares2021influence}
Soares L, de~Oliveira J, de~Oliveira M and Ferreira S 2021 {\em Physica A:
  Statistical Mechanics and its Applications\/} {\bf 573} 125931

\bibitem{ising1925beitrag}
Ising E 1925 {\em Zeitschrift f{\"u}r Physik\/} {\bf 31} 253--258

\bibitem{stauffer2007ising}
Stauffer D and Aharony A 2007 {\em Introduction to percolation theory\/} {\bf
  91} 46--49

\bibitem{newman1999monte}
Newman M~E~J and Barkema G~T 1999 {\em Monte Carlo methods in statistical
  physics\/} vol~1 (Oxford University Press)

\bibitem{kuperman2001small}
Kuperman M~N and Abramson G 2001 {\em Physical Review Letters\/} {\bf 86} 2909

\bibitem{snajd2005}
Snajd O and Slanina F 2006 {\em The European Physical Journal B\/} {\bf 50}
  93--97

\bibitem{sznajd2000opinion}
Sznajd-Weron K and Sznajd J 2000 {\em International Journal of Modern Physics
  C\/} {\bf 11} 1157--1165

\bibitem{lima2007nonequilibrium}
Lima F~W and Gon{\c{c}}alves S 2007 {\em Physical Review E\/} {\bf 76} 036117

\bibitem{rodrigues2005surviving}
Rodrigues F~A and da~F~Costa L 2005 {\em International Journal of Modern
  Physics C\/} {\bf 16} 1785--1792

\bibitem{liggett1985interacting}
Liggett T~M 1985 {\em Interacting particle systems\/} (Springer-Verlag)

\bibitem{baronchelli2006sharp}
Baronchelli A, Dall'Asta L and Barrat A 2006 {\em Journal of Statistical
  Mechanics: Theory and Experiment\/} {\bf 2006} P08014

\bibitem{baronchelli2006mean}
Baronchelli A and Dall’Asta L 2006 {\em Physica A: Statistical Mechanics and
  its Applications\/} {\bf 356} 294--299

\bibitem{baronchelli2006naming}
Baronchelli A and Loreto V 2006 {\em Journal of Statistical Mechanics: Theory
  and Experiment\/} {\bf 2006} P08001

\bibitem{deffuant2000mixing}
Deffuant G, Neau D, Amblard F and Weisbuch G 2000 {\em Advances in Complex
  Systems\/} {\bf 3} 87--98

\bibitem{hegselmann2002opinion}
Hegselmann R and Krause U 2002 {\em Journal of artificial societies and social
  simulation\/} {\bf 5} 2

\bibitem{malarz2009analysis}
Malarz K and Kulakowski K 2009 {\em Physical Review E\/} {\bf 80} 036105

\bibitem{pinheiro2015asymmetric}
Pinheiro F~L and Santos M~A 2015 {\em Physical Review E\/} {\bf 92} 012812

\bibitem{doniec2022consensus}
Doniec M, Lipiecki A and Sznajd-Weron K 2022 {\em Entropy\/} {\bf 24} 983

\bibitem{lipiecki2022polarization}
Lipiecki A and Sznajd-Weron K 2022 {\em Chaos, Solitons \& Fractals\/} {\bf
  165} 112809

\bibitem{krueger2017conformity}
Krueger T, Szwabi{\'n}ski J and Weron T 2017 {\em Entropy\/} {\bf 19} 371

\bibitem{soares2021modular}
Soares R~L and Fontanari J~F 2021 {\em Physical Review E\/} {\bf 103} 032312

\bibitem{fernandez2014effect}
Fern{'a}ndez-Gracia J, Egu{'\i}luz V~M and San~Miguel M 2014 {\em Physical
  Review E\/} {\bf 90} 012811

\bibitem{de2013effects}
De~Sanctis L~F, Gon{\c{c}}alves B and Pinto S~S 2013 {\em Physical Review E\/}
  {\bf 87} 052811

\bibitem{liu2020effect}
Liu Z, Wang J, Ren F and Wu X 2020 {\em EPL (Europhysics Letters)\/} {\bf 130}
  20001

\bibitem{brugnano2018role}
Brugnano G, Chat{'e} H and Manrubia S~C 2018 {\em Physical Review E\/} {\bf 97}
  062309

\bibitem{keeling2005networks}
Keeling M~J and Eames K~T 2005 {\em Journal of the Royal Society Interface\/}
  {\bf 2} 295--307

\bibitem{rodrigues2019machine}
Rodrigues F~A, Peron T, Connaughton C, Kurths J and Moreno Y 2019 {\em arXiv
  preprint arXiv:1910.00544\/}

\bibitem{pecora1998synchronization}
Pecora L~M and Carroll T~L 1998 {\em Physical review letters\/} {\bf 80} 2109

\bibitem{brooks2020model}
Brooks H~Z and Porter M~A 2020 {\em Physical Review Research\/} {\bf 2} 023041

\bibitem{moretti2013mean}
Moretti P, Liu S, Castellano C and Pastor-Satorras R 2013 {\em Journal of
  statistical physics\/} {\bf 151} 113--130

\bibitem{lorenz2007continuous}
Lorenz J 2007 {\em International Journal of Modern Physics C\/} {\bf 18}
  1819--1838

\bibitem{moretti2013griffiths}
Moretti P and Munoz M~A 2013 {\em Nature communications\/} {\bf 4} 1--6

\bibitem{guerra2004information}
Guerra R~A, Egu{\'\i}luz V~M and San~Miguel M 2004 {\em Physical Review E\/}
  {\bf 70} 046106

\bibitem{mobilia2003single}
Mobilia M 2003 {\em Physical Review Letters\/} {\bf 91} 028701

\bibitem{galam2002minority}
Galam S 2002 {\em The European Physical Journal B-Condensed Matter and Complex
  Systems\/} {\bf 25} 403--406

\bibitem{castellano2009nonlinear}
Castellano C, Mu{\~n}oz M~A and Pastor-Satorras R 2009 {\em Physical Review
  E\/} {\bf 80} 041129

\bibitem{realqvoter}
Jankowski R 2020 real-q-voter
  \url{https://github.com/robertjankowski/real-q-voter} accessed on: September
  6, 2023

\bibitem{erdHos1960evolution}
Erd{\H{o}}s P, R{\'e}nyi A {\em et~al.\/} 1960 {\em Publ. Math. Inst. Hung.
  Acad. Sci\/} {\bf 5} 17--60

\bibitem{barabasi1999emergence}
Barab{\'a}si A~L and Albert R 1999 {\em science\/} {\bf 286} 509--512

\bibitem{onody2004nonlinear}
Onody R~N and de~Castro P~A 2004 {\em Physica A: Statistical Mechanics and its
  Applications\/} {\bf 336} 491--502

\bibitem{lancichinetti2008benchmark}
Lancichinetti A, Fortunato S and Radicchi F 2008 {\em Physical review E\/} {\bf
  78} 046110

\bibitem{watts1998collective}
Watts D~J and Strogatz S~H 1998 {\em nature\/} {\bf 393} 440--442

\bibitem{waxman1988routing}
Waxman B~M 1988 {\em IEEE journal on selected areas in communications\/} {\bf
  6} 1617--1622

\bibitem{hagberg2014networkx}
Hagberg A~A, Swart P~J and Chult D~B 2014 Networkx
  \url{https://networkx.github.io/} version 1.9.1

\bibitem{freeman1979centrality}
Freeman L~C 1979 {\em Social Networks\/} {\bf 1} 215--239

\bibitem{freeman1977set}
Freeman L~C 1977 {\em Sociometry\/} {\bf 40} 35--41

\bibitem{newman2010networks}
Newman M~E 2010 {\em Networks: An Introduction\/} (Oxford University Press)

\bibitem{newman2018networks}
Newman M~E 2018 {\em Networks: An Introduction\/} (Oxford University Press)

\bibitem{stephenson1989rethinking}
Stephenson K and Zelen M 1989 {\em Social Networks\/} {\bf 11} 1--37

\bibitem{estrada2008communicability}
Estrada E and Hatano N 2008 {\em Physical Review E\/} {\bf 77} 036111

\bibitem{brandes2005modularity}
Brandes U, Delling D, Gaertler M, Goerke R, Hoefer M, Nikoloski Z and Wagner D
  2005 {\em IEEE Transactions on Knowledge and Data Engineering\/} {\bf 17}
  754--767

\bibitem{bonacich1987power}
Bonacich P 1987 {\em American Journal of Sociology\/} {\bf 92} 1170--1182

\bibitem{costa2007characterization}
Costa L~d~F, Rodrigues F~A, Travieso G and Villas~Boas P~R 2007 {\em Advances
  in physics\/} {\bf 56} 167--242

\bibitem{wainer2021nested}
Wainer J and Cawley G 2021 {\em Expert Systems with Applications\/} {\bf 182}
  115222

\bibitem{cawley2010over}
Cawley G~C and Talbot N~L 2010 {\em The Journal of Machine Learning Research\/}
  {\bf 11} 2079--2107

\bibitem{leaver2018fronto}
Leaver A~M, Wade B, Vasavada M, Hellemann G, Joshi S~H, Espinoza R and Narr K~L
  2018 {\em Frontiers in psychiatry\/} {\bf 9} 92

\bibitem{nakagawa2017coefficient}
Nakagawa S, Johnson P~C and Schielzeth H 2017 {\em Journal of the Royal Society
  Interface\/} {\bf 14} 20170213

\bibitem{spadon2019reconstructing}
Spadon G, Carvalho A~C~d, Rodrigues-Jr J~F and Alves L~G 2019 {\em Scientific
  reports\/} {\bf 9} 11801

\bibitem{centola2007complex}
Centola D and Macy M 2007 {\em American journal of Sociology\/} {\bf 113}
  702--734

\bibitem{sunstein2018republic}
Sunstein C 2018 {\em \# Republic: Divided democracy in the age of social
  media\/} (Princeton university press)

\bibitem{networkxerdosrenyidoc}
{NetworkX} 2023 Erdős-rényi graphs
  \url{https://networkx.org/documentation/stable/reference/generated/networkx.generators.random_graphs.erdos_renyi_graph.html}

\bibitem{igraphbarabasidoc}
{Igraph} 2023 Barabasi-albert model
  \url{https://igraph.org/python/api/latest/igraph._igraph.GraphBase.html#Barabasi}

\bibitem{networkxdoc}
{NetworkX} 2023 Lfr benchmark graph
  \url{https://networkx.org/documentation/stable/reference/generated/networkx.generators.community.LFR_benchmark_graph.html}

\bibitem{networkxwattsstrogatzdoc}
{NetworkX} 2023 Watts-strogatz graph
  \url{https://networkx.org/documentation/stable/reference/generated/networkx.generators.random_graphs.watts_strogatz_graph.html}

\bibitem{networkxwaxmandoc}
{NetworkX} 2023 Waxman graph
  \url{https://networkx.org/documentation/stable/reference/generated/networkx.generators.geometric.waxman_graph.html}

\bibitem{networkxpathgraphdoc}
{NetworkX} 2023 Path graph
  \url{https://networkx.org/documentation/stable/reference/generated/networkx.generators.classic.path_graph.html}

\bibitem{qvotergithub}
Paul~Kent A~P 2023 Qvml 2023 \url{https://github.com/kentwar/QVML\_2023/}

\end{thebibliography}

\end{document}